%% file: main.tex
\documentclass[twocolumn]{aastex631}
\submitjournal{ApJ}
\shorttitle{CHAMPSS System Overview}
\shortauthors{CHAMPSS Collaboration et al.}

\usepackage[T1]{fontenc}
\usepackage{enumitem}
\usepackage{multirow}
\usepackage{booktabs}
\newcommand{\dmu}{\,\textrm{pc}\,\textrm{cm}^{-3}}

\begin{document}

\title{CHIME All-sky Multiday Pulsar Stacking Search (CHAMPSS): System Overview and First Discoveries}

\collaboration{99}{CHAMPSS Collaboration}
\input{authors.tex}
\correspondingauthor{Robert Main, Lars K\"{u}nkel}
\email{robert.main@mcgill.ca, lkuenkel@phas.ubc.ca}

\begin{abstract}

We describe the CHIME All-sky Multiday Pulsar Stacking Search (CHAMPSS) project. 
This novel radio pulsar survey revisits the full Northern Sky daily, offering unprecedented opportunity to detect highly intermittent pulsars, as well as faint sources via long-term data stacking.
CHAMPSS uses the CHIME/FRB datastream, which consists of 1024 stationary beams streaming intensity data at $0.983$\,ms resolution, 16384 frequency channels across 400--800\,MHz, continuously being searched for single, dispersed bursts/pulses.  
In CHAMPSS, data from adjacent east-west beams are combined to form a grid of tracking beams, allowing longer exposures at fixed positions. 
These tracking beams are dedispersed to many trial dispersion measures (DM) to a maximum DM beyond the Milky Way's expected contribution, and Fourier transformed in time to form power spectra.  
Repeated observations are searched daily to find intermittent sources, and power spectra of the same sky positions are incoherently stacked, increasing sensitivity to faint persistent sources.
The $0.983$\,ms time resolution limits our sensitivity to millisecond pulsars; we have full sensitivity to pulsars with $P > 60\,$ms, with sensitivity gradually decreasing from $60$ ms to $2$\,ms as higher harmonics are beyond the Nyquist limit.
In a commissioning survey, data covering $\sim 1/16$ of the CHIME sky was processed and searched in quasi-realtime over two months, leading to the discovery of eleven new pulsars, each with $S_{600} > 0.1$\,mJy. 
When operating at scale, CHAMPSS will stack $>$1\,year of data along each sightline, reaching a sensitivity of $\lesssim 30\, \mu$Jy for all sightlines above a declination of $10^{\circ}$, and off of the Galactic plane.

\end{abstract}


\section{Introduction} \label{sec:intro}

There are $>$3700 pulsars discovered to date \citep{manchester+05}, enabling a wealth of physics and astrophysics.  
Relativistic binary systems revealed indirect evidence for gravitational waves \citep{taylor+89}, and continue to be a laboratory for the most precise tests of general relativity (e.g., \citealt{kramer+21}).  
Pulsar Timing Arrays, which combine the timing residuals of the most precisely timed pulsars, are starting to detect evidence for $\sim$\,nHz gravitational waves from supermassive black hole binaries \citep{nanograv23, epta23, ppta23}.  
By measuring pulsar masses \citep{demorest+10, fonseca+21} and radii \citep{miller+21}, the neutron star equation of state is constrained thereby providing insight into how matter behaves at extreme densities \citep{lattimer+07}.

Pulsar signals are also a powerful probe of intervening plasma.  
Pulses are dispersed, acquiring a frequency-dependent delay directly proportional to the total electron column along the line of sight.  
Moreover, pulses are often highly linearly polarized, allowing measures of the Faraday Rotation, probing intervening magnetic fields.  
Pulsars experience multipath propagation owing to small-scale variations in electron density, leading to scattering and, as pulsars are effectively point sources, interference effects between multiple deflected paths, known as `scintillation'. 
Measuring the above effects on many pulsar sightlines led to Galactic electron models \citep{ne2001, ymw16}, foreground maps of Galactic magnetism \citep{han+06}, and a holographic view of plasma substructure on $\lesssim 0.1$\,AU \citep{stinebring+22}. 

To study pulsars, first they must be discovered.  
Most pulsar surveys to date used traditional parabolic dishes.  
Due to the small field of view (FoV), the strategy is typically to gradually tile the sky with pointings, or to focus on regions where pulsars are \textit{a priori} expected to reside.  
The most obvious choice is to search in the Galactic Plane, although pulsars also reside in globular clusters ($>340$ to date\footnote{\url{https://www3.mpifr-bonn.mpg.de/staff/pfreire/GCpsr.html}}), supernova remnants \citep{staelin+68, large+68}, are often associated with $\gamma-$ray sources \citep{smith+23}, and can sometimes be seen as compact steep-spectrum radio continuum sources \citep{backer+82}.  
Both the Five-hundred-meter Aperture Spherical Telescope (FAST) and MeerKAT are performing a Galactic Plane survey \citep{han+25, padmanabh+23}, also done as part of the Parkes Multibeam Survey \citep{manchester+01}.

All-sky surveys such as the Green Bank Northern Celestial Cap Pulsar Survey \citep[GBNCC;][]{stovall+14} require much longer to survey the full sky, and are typically limited to one short exposure per pointing.
Low frequency telescopes which form beams digitally, such as the Low Frequency Array (LOFAR) and Murchison Widefield Array (MWA), can have a much larger FoV and survey their full visible sky more rapidly. 
The LOFAR Tied-Array All-Sky Survey (LOTASS) forms many tied-array beams in real-time within the LOFAR FoV (from just the inner dense core stations) at ${\sim}135\,$MHz, and searches them offline \citep{sanidas+19}. 
Similarly, the Southern-sky MWA Rapid Two-metre (SMART) pulsar survey utilizes the MWA FoV (in its compact configuration) at ${\sim}154$\,MHz to rapidly survey the southern sky for pulsars down to ${\sim}$2-3\,mJy with up to 80-minute dwell times, but at the cost of very large offline computation and storage footprints by virtue of processing the raw tile voltage data \citep{smart1,smart2}.

In this paper, we describe a new pulsar survey using the Canadian Hydrogen Intensity Mapping Experiment \citep[CHIME;][]{chime-overview}.  
The unique cylindrical nature of CHIME means it has a much larger field of view than a parabolic dish, and is an ideal survey instrument. 
CHIME has been transformative to the field of Fast Radio Bursts \citep{chime-frbcat1}, and the real-time system to find individual bursts has already led to the discovery of more than 80 pulsars\footnote{\url{https://www.chime-frb.ca/Galactic}} \citep{good+21, dong+23}.  
The CHIME All-sky Multiday Pulsar Stacking Survey (CHAMPSS), aims to use the large field of view of CHIME to carry out a daily full-sky Fast Fourier Transform (FFT) search for pulsars.
Power spectra from repeated observations of the same positions will be incoherently stacked \citep{vdklis+1989}, a method which has been successfully used to discover pulsars in globular clusters \citep{bruce+1993, paz+16, cadellano+18}.
Through the combination of daily searching and stacking of the full northern sky,
this survey will be deeper than any other full-sky survey to date, and will be sensitive to intermittent sources which could have by-chance been missed previously.  
This can occur if, e.g., pulsars are scintillating \citep{rickett90}, eclipsed \citep{fruchter+88, johnston+92}, precessing \citep{breton+08}, nulling \citep{backer70}, intrinsically intermittent \citep{lyne+10}, or simply if observations are corrupted by Radio Frequency Interference (RFI). 
CHAMPSS will additionally lead to pulsar detections on under-searched lines-of-sight, providing a better sampling of the Galactic electron structure.

The distribution of the paper is as follows:  In Section 2 we describe CHIME, as well as the FRB and Pulsar datastreams relevant for CHAMPSS. In Section 3 we discuss the pipeline, how power spectra are formed from the incoming data stream, and how candidates are then sifted and grouped.  In Section 4 we detail candidate verification through a phase-coherent search, and timing of new-found sources.  In Section 5 we describe our operations and realtime system for processing and analyzing data.  In Section 6 we describe our commissioning survey, first discoveries, and implications.  Section 7 details the current status of CHAMPSS, planned expansion and forecast for the full survey.

\section{CHIME Systems}

Located at the Dominion Radio Astrophysical Observatory (DRAO) near Penticton, British Columbia, CHIME is a radio telescope comprised of $4\times100$\,m parabolic cylinders, oriented north-south \citep{chime-overview}.  Each cylinder has 256 dual-polarization linear feeds, operating in the frequency range 400--800\,MHz.  CHIME operates as a drift-scan telescope, observing the full northern sky above declinations of $-10^{\circ}$, with an instantaneous field-of-view of $\gtrsim 200$ square degrees.

We briefly overview the two crucial datastreams used for CHAMPSS, namely CHIME/FRB which is the backbone of our search, and CHIME/Pulsar which aids in confirming candidates and timing newly discovered pulsars.

\subsection{CHIME/FRB}

The CHIME/FRB system is described in detail in \citet{chimefrb18}; here we summarize the first steps, before CHAMPSS accesses the data.  

The CHIME correlator has two stages.  First, the F-engine channelizes the incoming data of 1024 dual-polarization receivers into 1024 channels each, resulting in time sampling of 2.56\,$\mu$s of the channelized data.  The X-engine is a GPU correlator which forms 1024 stationary beams on the sky (256 north-south $\times$ 4 east-west, \citealt{ng+17}).  Known as `L0', this stage also performs an additional FFT of each 128 samples both to upchannelize the data and downsample in time. After an additional factor of 8 averaging in frequency channels, and averaging every 3 successive time samples, the output is a datastream of $0.98304$\,ms, 16384\,channels per formed beam, as 8-bit intensity data.

The rest of the CHIME/FRB pipeline consists of four layers dubbed `L1' through `L4'.  The L1 stage of the pipeline performs initial RFI rejection, doing sigma-clipping from a series of polynomial and spline detrendings of the data in time and frequency (details in \citealt{chimefrb18}).  It is at this stage after the L1 RFI masking that CHAMPSS taps into the CHIME/FRB datastream, before the dedispersion and subsequent single-burst searches.  While CHAMPSS also needs to perform a dedispersion step, we require additional filtering of periodic sources of RFI before the transform, described in Section \ref{sec:fdmt}.

\vspace{1cm}
\subsection{CHIME/Pulsar}

Running in parallel, CHIME/Pulsar can simultaneously form up to 10 tracking beams on the sky.  The system observes sources with a probabilistic scheduler, where each source has a tunable priority ranking \citep{chimepulsar21}.  CHIME/Pulsar operates on complex baseband data with 1024\,channels, 2.56\,$\mu$s resolution, and can be used to produce either fold-mode or search-mode data.  Fold-mode data are coherently dedispersed to a specific DM and folded according to an ephemeris, forming an archive; a data cube with dimensions of subintegration, polarization, frequency, and pulse phase.  Search-mode data can be coherently dedispersed to a specific DM, with the data produced being intensity as a function of time and frequency; this can later be search for individual pulses or folded.

The tracking beam, finer time resolution, and coherent dedispersion all make CHIME/Pulsar comparatively more sensitive for targeted observations, while the wide field-of-view of CHIME/FRB is better for searches.  For CHAMPSS, CHIME/Pulsar is used to help confirm and time candidates, as described in \ref{sec:timing}.

\section{CHAMPSS Pipeline: Acquisition, Reduction and Search}
\label{sec:pipeline}

\begin{figure}[ht]
    \centering
    \includegraphics[width=1.0\columnwidth, trim={6cm 1.5cm 4cm 1.5cm}, clip]{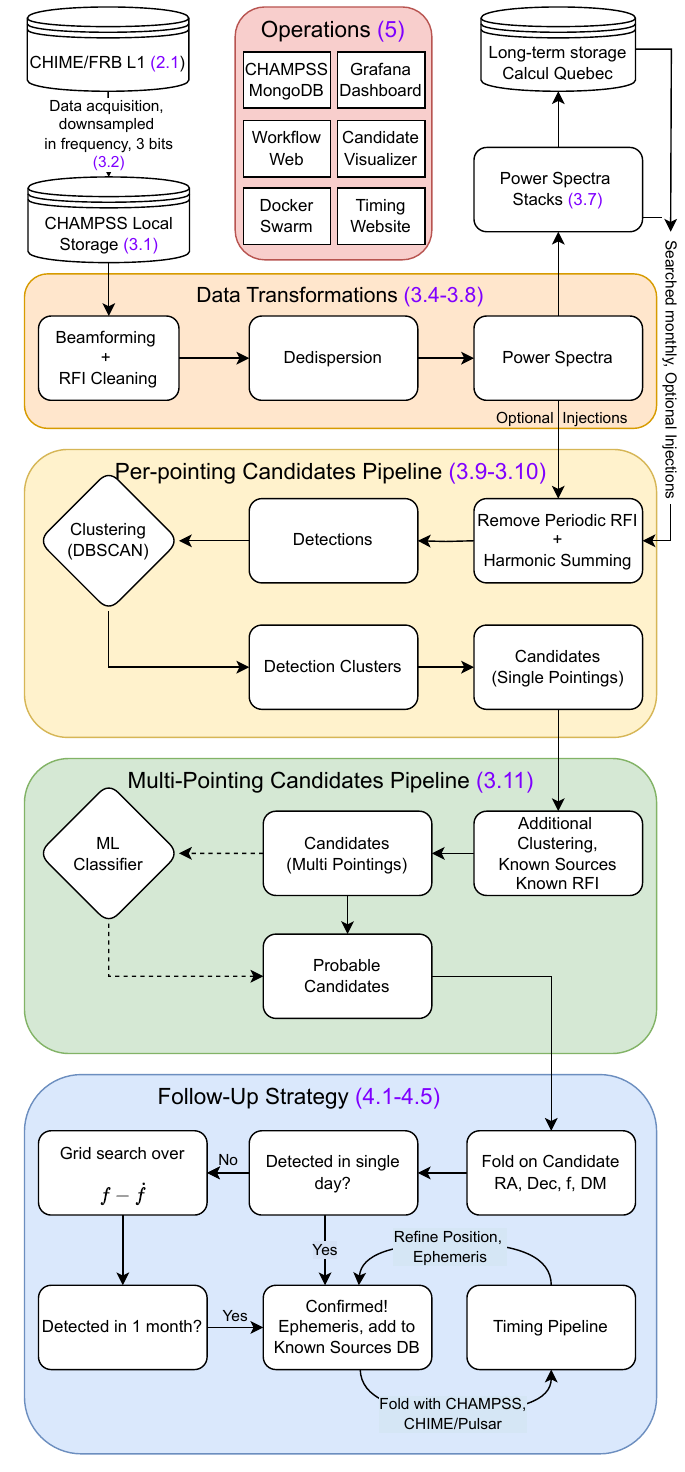}
    \caption{Simplified flowchart of the CHAMPSS pipeline, as described in Sections \ref{sec:pipeline}, \ref{sec:followup}. The dotted arrow to the Machine Learning candidate classifier indicates that it is a planned component of our pipeline, which was not used in our commissioning survey. }
    \label{fig:flowchart}
\end{figure}

In this Section, we describe the different stages of the pipeline, starting from the CHIME/FRB datastream, resulting in power spectra (daily and stacked) and clustered/filtered candidates.  A schematic flowchart is shown in Figure \ref{fig:flowchart}.  

Our regularly updated codebase is located at \url{https://github.com/chime-sps/champss_software}, and the sub-repositories within.

\subsection{Computing Setup}
\label{sec:computing}
Installed in 2022, we have a commissioning cluster for CHAMPSS located at the CHIME site. There are two compute nodes, each with 128 logical cores (64 physical cores, utilizing AMD's Simultaneous Multithreading), and 256\,GB of RAM.  These are connected via 10\,GbE link to an archiver node with 16$\times$14\,TB hard drives, for a total 208\,TB storage in a Zettabyte File System (ZFS) Pool. This cluster is connected to the L1 nodes of CHIME/FRB via a 40\,GbE link. Our storage and compute capacities will be greatly expanded to increase the sky capacity of the search, details will be provided in future works.

The aforementioned cluster is used to process incoming data in quasi-realtime (described in the following sections), as the data rate is sufficiently high that it cannot be copied off site rapidly enough.  We do however copy long-term data products to a supercomputing cluster `Narval', owned and operated by Calcul Québec, where we have a 2.2\,PB storage allocation.  The timing pipeline (Section \ref{sec:timing}) is run there, as will be the search and long-term storage of power spectrum stacks (Section \ref{sec:powerstack}).

\subsection{CHAMPSS Data Description and Acquisition}
\label{sec:data}

As mentioned above, CHAMPSS begins with the CHIME/FRB datastream, which has $16384$ channels, and 0.98306\,ms time resolution. This corresponds to $\sim 1.5$\,PB/day, for the full datastream from all 1024 beams.  However, the data rate for CHAMPSS can be greatly reduced; we can record with fewer channels in most parts of the sky as we only need to be sensitive up to the DM from the Milky Way, not for extragalactic sources. As described in Section \ref{sec:pointingmap}, we base our pointing map on known Galactic electron models, with conservative errors on the max DM allowing for model uncertainties, halo contribution, and even Local Group galaxies.  Additionally, while the data are stored in 8-bit integers, less dynamic range is needed for faint, periodic pulsar signals. 
The $256\times4$ grid of formed beams searched by CHIME/FRB extends from declinations of $\delta=-10^{\circ}$, up to the North pole at $\delta=90^{\circ}$, and beyond the pole down to $\delta\approx 75^{\circ}$.  We ignore the "lower transit" beams beyond the pole because they are less sensitive due to the lower effective collecting area towards the horizon, and since they overlap declinations covered by the lower beam rows.  This leaves 224 of the 256 beams.

The L1 nodes on CHIME/FRB have a continuous data buffer for each of the 1024 beams.  To record the data for a specific beam we use a Remote Procedure Call (RPC) from our archiver, mapped to the corresponding node.  The RPC requests the desired reduction in frequency resolution and bit depth, where the frequency downsampling is governed by the optimal number of channels to reach the maximum DM along the given sightline (described in Section \ref{sec:pointingmap}).   
In blocks of 1024 time samples (${\sim}1\,$s), the data are downsampled on the L1 nodes, and saved per-channel $\nu_i$ as $(I(\nu_i, t) - \mu(I(\nu_i)) ) / \sigma(I(\nu_i))$, along with a header containing the starting timestamp, the RFI mask, along with the per-channel mean $\mu(I(\nu_i))$ and standard deviation $\sigma(I(\nu_i))$ used in the normalization above.  The downsampled data are saved with 3-bit resolution, following Huffman coding, which is a scheme to compress data with minimal loss of information \citep{huffman52}. The encoding has 5 levels for data values, and gives an estimated loss of information of $\sim 5\%$, when the S/N per sample is low. With both the channel reduction and bit reduction, the daily data rate for CHAMPSS is reduced to 52\,TB.  We note that at the time of writing, a bug was identified where normalizations can jump at the $1\%$ level between blocks in downsampled data, leading to increased red noise at $f\lesssim 1$\,Hz -- this reduced the sensitivity of our pilot survey but will be fixed for any further data taken.

\subsection{Pointing Map and DM limits}
\label{sec:pointingmap}

Unlike CHIME/FRB, which is content with static beams, we wish to grid the sky into pointed observations at a fixed right ascension ($\alpha$) and declination ($\delta$). The separation in $\delta$ matches the CHIME/FRB beams by design, and we choose the separation in $\alpha$ between adjacent pointings such that they are $\Delta \alpha = 0.32/\cos(\delta)$ degrees apart.  For the full CHIME sky, this amounts to 165537 independent pointings.  The pointing durations differ based on the declination, due to the transit nature of CHIME. 

The maximum expected DM for a Galactic pulsar is highly position dependent, with the total model DM ranging from $\lesssim 100\dmu$ far off the Galactic Plane, and $>4000\dmu$ in the direction of the Galactic Centre for a pulsar at the far edge of the MW. For sightlines with low maximum DM, we can reduce the channelization without suffering additional DM smearing; i.e. we wish to choose an optimal channelization to minimize CPU time and storage, without introducing DM smearing larger than our sampling time.  

We form a Galactic DM map using the two most used Galactic electron models, NE2001 \citep{ne2001} and YMW16 \citep{ymw16}, as follows:
\begin{itemize}
    \item For each $\alpha, \delta$ we compute the maximum DM along the line of sight as $\rm{DM}(\alpha, \delta, D=50\,{\rm kpc})$, taking the maximum of NE2001 or YMW16 as $\rm{DM}_{\rm max}(\alpha, \delta)$.
    \item For Galactic latitudes $|g_b| > 15^\circ$, $\rm{DM}_{\rm search} = 2 \rm{DM}_{\rm max} + 50 \dmu$.
    \item For Galactic latitudes $|g_b| < 15^\circ$, $\rm{DM}_{\rm search} = \rm{DM}_{\rm max} \times \textrm{exp}(0.0313\times |g_b| + 0.223)$, a heuristic function which rises from 1.25 at $|g_b|=0^\circ$ to $2.0$ at $|g_b| = 15^\circ$.
    \item Along sightlines to M31 and M33, we add an additional $400\dmu$.
\end{itemize}
Galactic electron models are less constrained at high Galactic latitudes, since there are fewer pulsars than in the plane. NE2001 predicts higher values of $\rm{DM}_{\rm max}$ than YMW16 at Galactic latitudes $|g_b| < 2^{\circ}$, and vice versa, with fractional differences of $\rm{DM}_{\rm max}$ between the two models in excess of $50\%$ along some sightlines (for a comparison, see \citealt{price+21}). By taking the conservative largest expected DM of either model, we expect few Galactic pulsars to be beyond our DM search limits, with the possible exception of pulsars lying behind HII regions (\citealt{ocker+24}, and see Section \ref{sec:cygnus}).
Aside from the Milky Way, M31 (Andromeda) and M33 (Triangulum) are the two most massive galaxies in the Local Group, at $D \approx 0.8\,$Mpc, $D \approx 0.9$\,Mpc, respectively.
 Their expected DM distributions have been modeled in \citet{ocker+22}, with the model reaching a maximum of $\rm{DM}\approx 400\dmu$ adopted above.

\begin{figure*}
    \centering
    \includegraphics[width=0.95\textwidth]{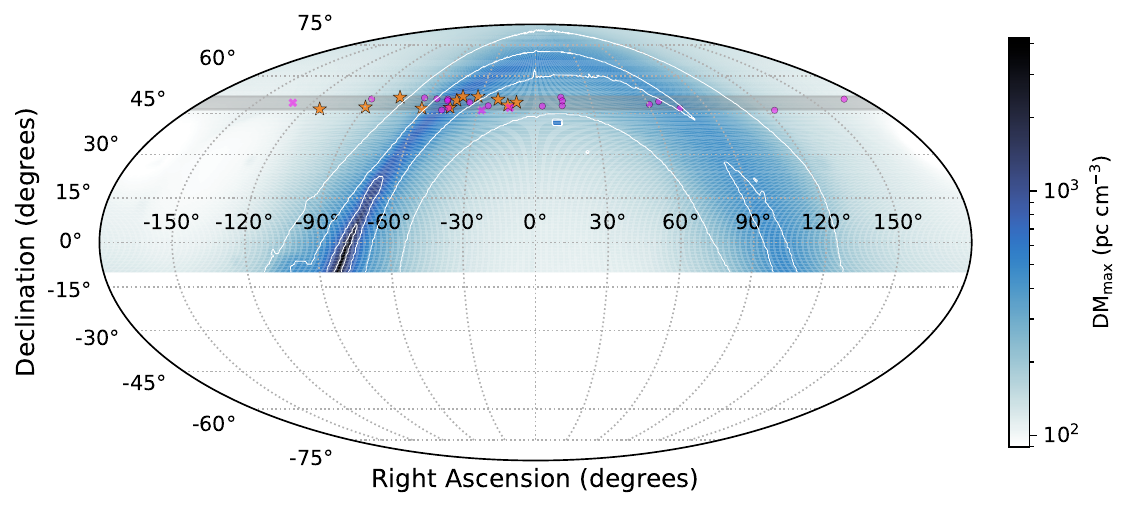}
    \caption{ CHAMPSS pointing map.  The color bar represents the maximum DM that we search to along a given sightline in units of pc cm$^{-3}$.  The contours indicate increasing number of channels in powers of 2, representing 5 tiers from 1024 to 16384. The grey shaded region denotes our commissioning survey, the orange stars denote the newly discovered pulsars, and the magenta points show known pulsars which our survey detected.}
    \label{fig:pointing-map}
\end{figure*}

\begin{figure}
    \centering
    \includegraphics[width=1.0\columnwidth]{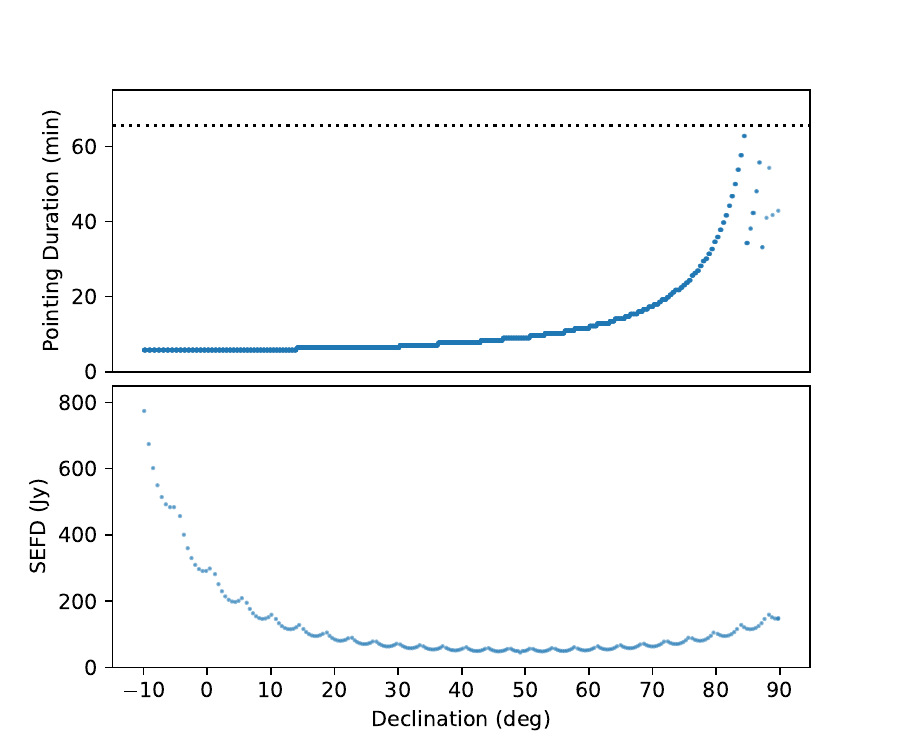}
    \caption{Duration (top) and sensitivity (bottom) of pointings as a function of declination. Pointings at high-dec diverge in time as sources spend longer in the beam; pointings longer than $2^{22}$\,samples $\approx 68.7\,$min are split, for memory limitations. }
    \label{fig:sens-duration-dec}
\end{figure}

We use the `DDplan' utility from \texttt{PRESTO} \citep{presto} to determine the optimal channelization given $\rm{DM}_{\rm search}(\alpha, \delta)$, and round up to the nearest power of 2.
Our pointing map is illustrated in Figure \ref{fig:pointing-map}. The duration and sensitivity as a function of declination are shown in Figure \ref{fig:sens-duration-dec}.  The sensitivity plot is illustrative for the cold sky, using the temperature and gain values outlined in \ref{sec:sensitivity}, and the frequency-averaged CHIME beam model\footnote{\url{https://chime-frb-open-data.github.io/beam-model/}} for the relative sensitivity across declination.

\subsection{Beamforming and Initial RFI Cleaning}

The CHAMPSS pipeline works first by providing a date and the RA and Dec of one of the pointings in our pointing maps. 
The pipeline then tests whether data were recorded during the time corresponding to this pointing. 
If data are present on disk, then a `beamforming' step is performed; each pointing in our pointing map is traversed by four static beams of CHIME/FRB, which can be combined to a quasi-tracking beam at fixed position. 
A combined time series of the data is formed by appending the data from these four beams based on when each beam is closest to the chosen pointing. 
This creates a single $I(\nu, t)$ data product, with corresponding RFI mask provided by the CHIME/FRB system.  

For further RFI cleaning, a known bad channel list which currently masks 21.6\% of channels is incorporated into the RFI mask and the remaining channels are passed through a filter based on the generalised spectral kurtosis estimator \citep{gsk1,gsk2,ngh16,nhm+16}.
This filtering step is notionally best suited to identify narrow-band, dynamic, and impulsive RFI. 
Each frequency channel is treated independently, on a timescale of 1024 time samples (1006.63296\,ms). 
The estimator statistic threshold ($\approx 25\sigma$ equivalent) used to flag additional data is set by the underlying data statistics and the data chunk size, and assumes that the channel data statistics are Gaussian without the presence of RFI.

For each segment corresponding to one input beam the values marked by the RFI mask are replaced by the median of the full segment. Linear trends from the data are then subtracted on scales corresponding to 32768 time samples\footnote{Using \texttt{scipy.signal.detrend} \citep{2020SciPy-NMeth}.}. The previously computed median of the segment is then re-added to the full segment. Afterwards the data are rescaled so that the data of each channel has the same median across all beam segments.

As a final step before dedispersion all frequency channels where more than 75\% of the samples are masked 
are set to 0. All other channels are normalised to have a mean of 0 and a standard deviation of 1.
Ultimately our individual RFI cleaning steps (the mask from the CHIME/FRB system, our known bad channel mask, the filter based on generalised spectral kurtosis estimator and the masking of largely masked channels), result in a median mask fraction of $\approx 55\%$.

\subsection{Dedispersion}
\label{sec:fdmt}

We dedisperse each pointing to many DM trials using the Fast Dispersion Measure Transform (FDMT) introduced by \citet{zackay+17}, an algorithm which dedisperses in $2N_{\nu} N_{t} + N_{t} N_{d} \log_{2}(N_{\nu})$ time, where $N_t, N_\nu, N_d$ are the number of time samples, frequency channels, and DM trials, respectively.  
The maximum DM for a given pointing is given by our CHAMPSS pointing map, and the DM spacing is set by the input time resolution, where the increment in DM corresponds to a delay of one time sample across the band.  
For storage and computational constraints, we calculate only every two DM trials, resulting in spacing of $\Delta \rm{DM} \approx 0.1\dmu$.
We use an implementation of FDMT in a  parallelized CPU code written in Python with C++ bindings\footnote{\url{https://github.com/pravirkr/dmt}}.  
After FDMT, the data product is then $I(\rm{DM}, t)$ per pointing.

\subsection{Power Spectra}
\label{sec:powerspectra}

For each of the dedispersed time series the power spectrum is computed after padding them to a length of $2^{20} = 1048576$ samples, resulting in a data product $I(\rm{DM}, f)$. 

In order to add power spectra from different days, we need to perform a barycentric correction. 
To achieve this in the power spectra domain, we apply the barycentric velocity correction to the sampling time of the observation, calculate the corresponding frequency bins, then interpolate (using nearest neighbor interpolation) to the frequency bins from the unaltered topocentric sampling time, which will be constant in time for each pointing.

After barycentric correction, we apply a red noise correction method from \texttt{PRESTO} to each of the power spectra. 
In this method a logarithmically increasing window is used at low frequencies to compute the local median. 
A linear fit to these windowed median values then provides a local red-noise estimate for each frequency bin, which is divided into the data value.

Two schemes for RFI suppression are used in the power spectrum domain. 
The first is a list of persistent `birdies' that show up regularly in our observations, and the second is a method to dynamically detect strong, unwanted RFI signals in each observation.
RFI signals will usually show up in many pointings across the sky while pulsars are more localised. This difference is used by the dynamic scheme;
we use the power spectrum I(\rm{DM}=0, f) (which we call the `zero DM power spectrum' throughout), find strong peaks ($>5\sigma$) and then add their frequency bins to the dynamic mask. 
In order to decide whether weaker peaks (2-5$\sigma$) should be removed, we compare them with the peaks of other nearby observations.
If a frequency bin is marked in more than 50\% of the compared pointings then it is added to the birdie list. 
To enable this comparison we store the birdie peaks in our observation database (see Section \ref{sec:database}) which allows other pipeline processes to access them. 
Since neighboring observations are needed for proper RFI removal, this necessitates that the available observations are processed in an order which guarantees that at least some nearby observations have been processed already. 
All frequency bins marked by the dynamic and static birdie filters are set to zero. 
These filters will mask a few percent of all frequency bins depending on the RFI situation.
The information about which frequency bins are masked is stored in order to use it in the search process. 

\subsection{Power Spectrum Stack}
\label{sec:powerstack}

\begin{figure}[ht]
    \centering
    \includegraphics[width=0.49\textwidth]{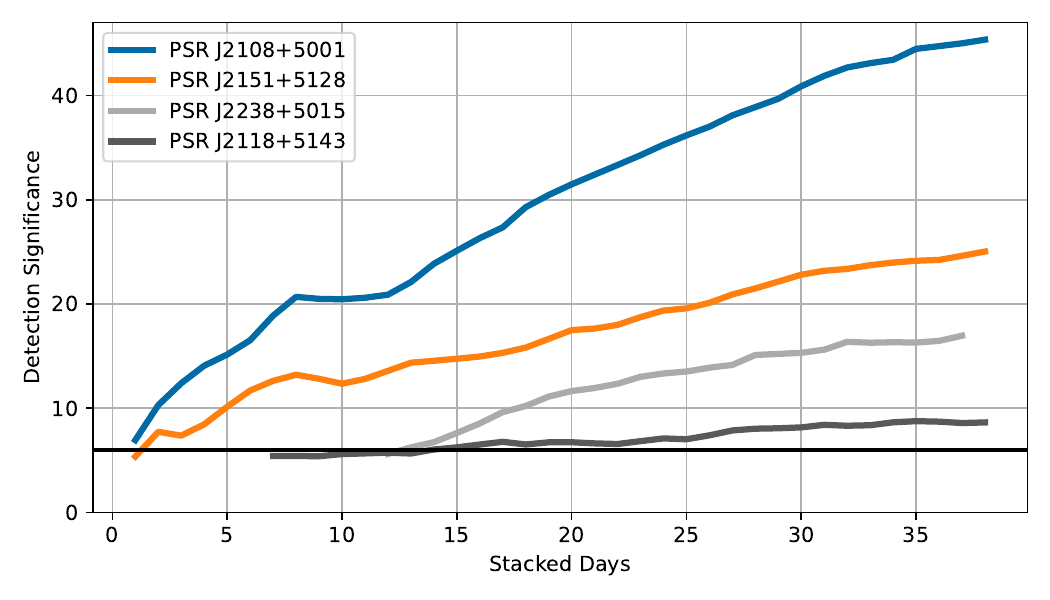}
    \caption{Candidate significance for four of our new pulsars as a function of number of days stacked. The horizontal black line shows the threshold of $6\sigma$ which we currently use during the stack searching process. In this run a threshold of $5\sigma$ was used and the stack was searched after each day which is not the case in our normal processing scheme.}
    \label{fig:sigma-dev}
\end{figure}

In order to gain sensitivity to faint pulsars we 
incoherently sum the power spectra from multiple days of observations of the same pointing. 
The effectiveness of this technique in allowing us to find pulsars that otherwise would have been missed in single observations is illustrated in Figure \ref{fig:sigma-dev}.
Retaining all power spectra from all observations on disk would not be possible for us due to storage constraints, which necessitates us summing the spectra from different days.

For each new daily observation, the power spectra are created as described above and first individually searched as outlined in the following sections. 
We compute quality metrics to determine whether a new set of power spectra should be added to the existing stack that is saved on disk. This filters out power spectra that may worsen our ability to find new pulsars in the stack due to them having unforeseen processing errors or strong RFI which has not been fully removed. For one of our quality metrics we perform the Kolmogorov-Smirnov test \citep{kolmogorov_1951} on the zero DM power spectrum to test whether it follows the $\chi^2$ distribution expected for our power spectrum stack \citep{vdklis+1989}.
We also compute the number of outliers in a $\chi^2$ distribution and compare it with the observed number of outliers in the zero DM spectrum.
Another quality metric that we use is the number of `detections' (see Section \ref{sec:detections}) that result from our search process, which will increase in the presence of a pulsar or RFI.
We expect strong RFI to increase the number of detections much more than any new pulsars we expect to find.

Once these quality metrics are computed, we compare them with static and dynamic thresholds. If a quality metric lies outside the maximum of these two thresholds, the power spectra are not added to the power spectrum stack. 
The static thresholds are manually set to adequate values to filter out bad power spectra. The dynamic thresholds are computed for each individual pointing; the median and median absolute deviation (MAD) of the previous values for the quality metric are computed, and dynamic threshold set to the median + $3\times$MAD.
The dynamic threshold allows us to use these quality metrics even if they are consistently higher than expected due to the presence of strong pulsars.

In this study we only employed one power spectrum stack for each pointing. In future CHAMPSS work where we will create much deeper stacks, we intend to save two different stacks on disk: a monthly stack which contains only relatively recent data and a cumulative stack which contains all previous data. The monthly stack allows a second pass of quality control to prevent the cumulative stack from containing too much RFI and may increase our sensitivity to intermittent pulsars. 
One set of power spectra for the full sky is roughly 400 TB.

\subsection{Injections}
\label{sec:injections}
In order to constrain the reliability of our pipeline, we have the capability to inject fake pulsar signals directly into the power spectrum. 
For this purpose, we use analytic templates derived from real pulse profiles from the MeerKAT Thousand Pulsar Array~\citep{posselt23} but vary injected significance, frequency, and DM.
We will inject
across all pointings to constrain the completeness of our pipeline,
and are tracking red noise across all pointings, 
to map it as a function of time and beam.

The injections only interact with the local version of the stack used for a given run of the pipeline, and so do not modify the database. 
The injection design and subsequent results on the transmissivity of the pipeline will be detailed in a future paper.

\subsection{Detections}
\label{sec:detections}

We follow \texttt{PRESTO}'s methodology for searching a power spectrum, summing harmonics within a power spectrum and evaluating the significance of any particular power.
For a time series of purely Gaussian noise the powers will follow a $\chi^2$ distribution with $2$ degrees of freedom, as already mentioned in Section \ref{sec:powerstack}. If $m$ such powers are summed together, whether in the harmonic summing procedure or power spectrum stacking, the result will follow a $\chi^2$ distribution with $2m$ degrees of freedom, $\chi^2_{2m}$ \citep{vdklis+1989,rem02}. To determine the significance of a power we therefore find the probability of this power occurring by chance in a $\chi^2_{2m}$ distribution, via the cumulative distribution, and convert this to a Gaussian-equivalent sigma.

By stacking power spectra we reach larger values of $m$ than are typically encountered in a pulsar search. The cumulative distribution of $\chi^2_{2m}$ is given by
\begin{equation}
F(P;\chi^2_{2m}) = \frac{\gamma(m, \frac{P}{2})}{\Gamma(m)}
\end{equation}
for a power $P$, $\gamma$ being the lower incomplete gamma function and $\Gamma$ the gamma function. In earlier stages of development, high values of $m$ led to overflow errors in the calculation of $\gamma$. To resolve this we use an algorithm specifically designed to avoid such errors \citep{abergelAlgorithm1006Fast2020}.

This search is performed for the power spectra at each trial DM, and at different harmonic sums
(1, 2, 4, 8, 16, and 32). As an example, for a harmonic sum with 4 harmonics at frequency $f$ the bins in the power spectrum corresponding to $f$, $2f$, $3f$, and $4f$ are summed together. Our significance threshold for the search is currently 5 and 6 Gaussian-equivalent $\sigma$ for searched daily observations and stacks respectively. Any points above this significance threshold are termed ``detections''. 
Detections in a single harmonic sum search of a single DM trial that are closer than $1.1 \times$
the frequency resolution of the power spectrum, are grouped together and only the most significant detection is kept.
Each detection contains information about its DM, frequency, Gaussian-equivalent $\sigma$, and the indices, frequencies, and powers of the individual bins in the power spectrum summed to produce the detection.

During the search step, we can also filter out known pulsars. This is performed by checking if a known pulsar has previously been identified in this pointing, using the known source sifter (see Section \ref{sec:multipointing}), and comparing the previously observed sigma with a given threshold. When looking for previously detected known pulsars, one can choose to look only at candidates resulting from single day spectra or from stacked spectra.
If the threshold is surpassed, the pulsar is removed by masking all frequency bins where the pulsar and its harmonics are expected.
This prevents strong pulsars from flooding the pipeline with detections, which would drastically worsen clustering performance (as described in the following Section). As the strength of the observed pulsars continuously increases as more days are stacked, this step is crucial when stacking more deeply.

\subsection{Clustering}

In a classic pulsar search, and earlier in the development of the pipeline, detections would be clustered in DM-frequency space to group detections from the same source together. The groups would then be assessed to determine if they were harmonically related, based on the most-significant frequency within the group, and filtered by this process. Detections at many harmonics of the fundamental frequency are expected for both RFI and pulsars, so this is a key stage in thinning the number of clusters to be made into candidates. 

RFI peaks are often broad in frequency, meaning the spread of frequencies in a group of detections resulting from RFI may be quite large. When assessing for harmonic relation, only using the most-significant frequency within a cluster led to many RFI clusters not meeting that criteria. 
As the vast majority of detections are due to RFI this was not ideal, and we implemented a system which clusters the detections in DM, frequency, and harmonic relation simultaneously.
This is achieved via scikit-learn's DBSCAN \footnote{\url{https://scikit-learn.org/stable/modules/generated/sklearn.cluster.DBSCAN.html}} algorithm using a custom metric which allows us to not only to identify clusters that are close in DM and frequency but also to identify harmonically related signals.
There is no simple transformation to convert harmonic relation into a nearest neighbors problem, and calculating this for each pair of detections would be prohibitively computationally expensive. 
Therefore, steps are taken to reduce the number of harmonic-relation calculations required. 
 
\begin{figure}[ht]
    \centering
    \includegraphics[width=0.49\textwidth]{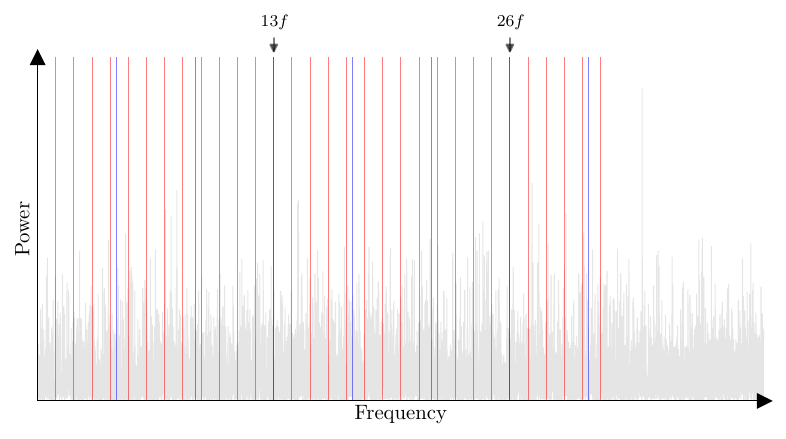}
    \caption{Example showing bins in a power spectrum shared by a 32 harmonic sum for a detection at $f$ (highlighted in red) and a 8 harmonic sum for a detection at $13f/3$ (blue).}
\label{fig:harmsum_example}
\end{figure}

The first step is to reduce the number of detections. 
If multiple detections occur at the same frequency and DM for different harmonic sums, only the most significant is kept. We also emulate \texttt{PRESTO} and eliminate any detections which are dominated by individual strong harmonics above the fundamental frequency of the detection, e.g. if a detection at frequency $f$ is dominated by a power at $7f$. 
The second is to identify detections which share the same frequency in order to avoid duplicate calculations. 
The last step uses the information about which bins were summed in the power spectrum. Most harmonically-related detections will share some power spectrum bins in their sum; for example a 8 harmonic sum to produce a detection at $\frac{13}{3}f$ will sum bin power spectrum bins
$$\frac{13f}{3}, \frac{26f}{3}, 13f, \frac{52f}{3}, \frac{65f}{3}, 26f, \frac{91f}{3}, \frac{104f}{3}$$
and clearly the $13f$ and $26f$ bins will be shared by a $f$ detection which summed 32 harmonics. An example of this is also shown in Figure \ref{fig:harmsum_example}. This property is utilized to form groups of detections which share power spectrum bins. Then, only detections within the same group are evaluated for harmonic relations.

Before the actual harmonic relations are calculated, an additional step of RFI filtering can be performed. 
This filtering step also uses DBSCAN but only utilises the frequency and DM as input using the detections that were not filtered out in the previous step. 
The DBSCAN run results in a list of detections that are dense in frequency and DM. 
Clusters that have their strongest $\sigma$ below a DM of 2$\dmu$ are filtered. 
Additionally, all clusters which have a mean frequency within the frequency span of those low-DM clusters can also be filtered out. 
This allows filtering of RFI signals which appear along a wide range of DM values. 
This step was added in October 2024 and not present when the pulsars in this paper were discovered.

During the processing of the results shown in this paper, we used two different approaches to create the final metric used in the DBSCAN clustering. 
In the first approach, the metric is computed based on the Euclidean distance in frequency and DM, then lowered if two detections are harmonically related. 
The originally calculated Euclidean distances is multiplied by a factor $k$ which is given by 
\begin{equation}
k = 1 - \frac{|B_1\cap B_2|}{\min(|B_1|, |B_2|)}
\end{equation}
where $B_1$ and $B_2$ are the frequency bins in the compared detections, $B_1 \cap B_2$ denoting the frequency bins that are in common between those detections and the $||$ operator denotes counting the number of frequency bins.
This was the clustering scheme used when our first pulsars were found.

In September 2024 we changed to a new scheme for the DBSCAN metric which will be used in future searches. 
In the new scheme, we not only compare the overlapping bins between detections but also use power in those frequency bins to reduce accidental clustering. 
In order to compare two detections which share frequency bins, we use the metric
\begin{equation}
M = 1- \max\left({\frac{P_{1\cap 2}}{P_1}, \frac{ P_{2\cap 1}}{P_2}}\right) + \Delta \text{DM}.
\end{equation}
$P_{1\cap 2}$ is the sum of the powers in the first detection which share frequency bins with the second detection. This is then divided by $P_1$, the total power in the first detection, to give the fraction of detection 1's power which lies in the shared bins, $\frac{P_{1\cap 2}}{P_1}$. Similarly $\frac{P_{2\cap 1}}{P_2}$ is the fraction of detection 2's power which lies in the shared bins.
$\Delta \text{DM}$ is the DM difference between the detections. This metric allows for better clustering because if a
signal is detected at a harmonic frequency, most of the power will still be contained in the bins shared with a detection at the fundamental frequency. 
In that case, the metric becomes very small and the detections are clustered. During this change, we also started using \texttt{scipy}'s sparse arrays and only computed the metrics for detections which are close in DM. This allows us to save on memory, which would otherwise limit the number of detections which can be clustered.

At the end of the clustering process, the identified clusters are then passed on to the candidate creation process.

\subsection{Candidate Creation}

\begin{figure}[ht]
    \centering
    \includegraphics[width=0.49\textwidth]{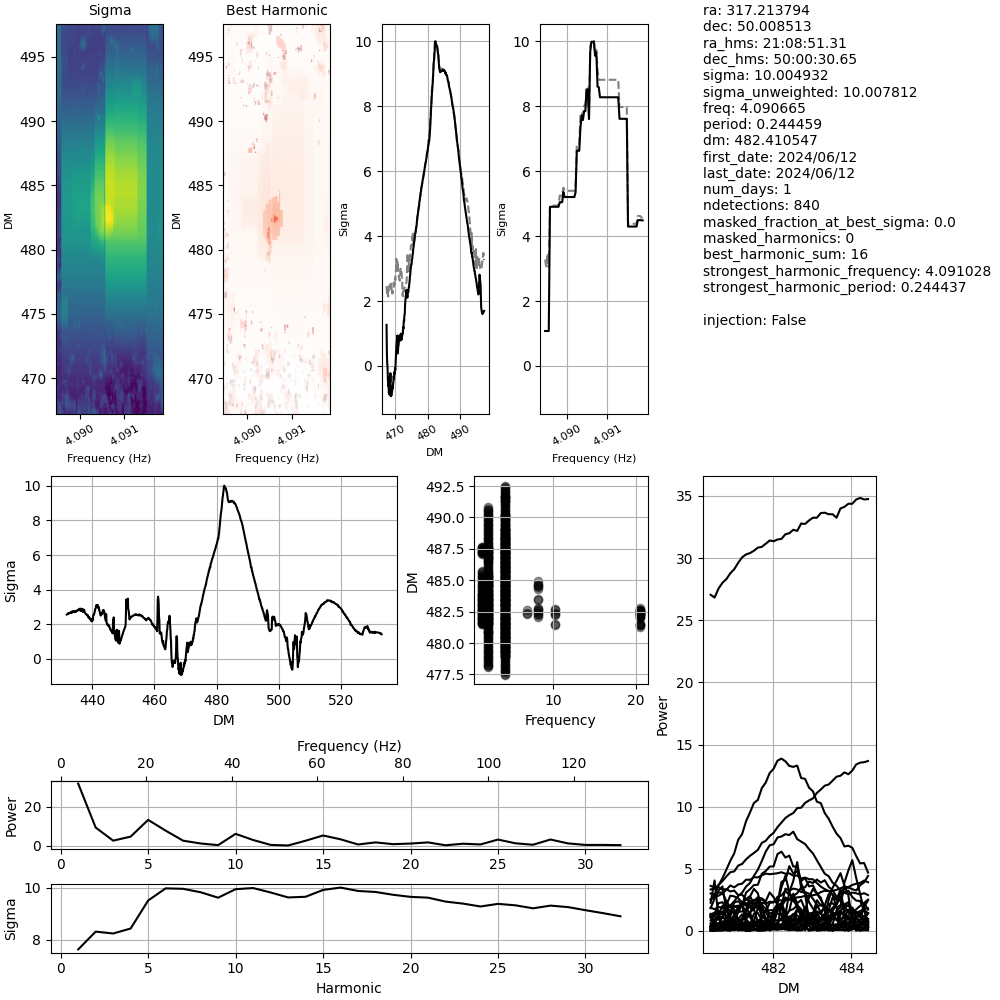}
    \caption{Candidate plot for a detection of our new pulsar PSR~J2108+5001 in a single observation. The various diagnostic plots show how the signal develops as a function of frequency and DM. It also shows the signal strength at various harmonic frequencies. The clustered detection which make up this candidate also contains detections at subharmonics which result from the harmonic summing.}
\label{fig:single_pointing_candidate}
\end{figure}

After the detections have been clustered, additional diagnostics are computed in order to save a rich representation of the detected signal on disk and create a candidate plot, example shown in Figure \ref{fig:single_pointing_candidate}. The diagnostics show how signals compare at nearby frequencies, nearby DMs and different harmonics which helps us distinguish between real pulsars and RFI candidates.

\label{sec:multipointing}

\begin{figure*}
    \centering
    \includegraphics[width=0.49\textwidth]{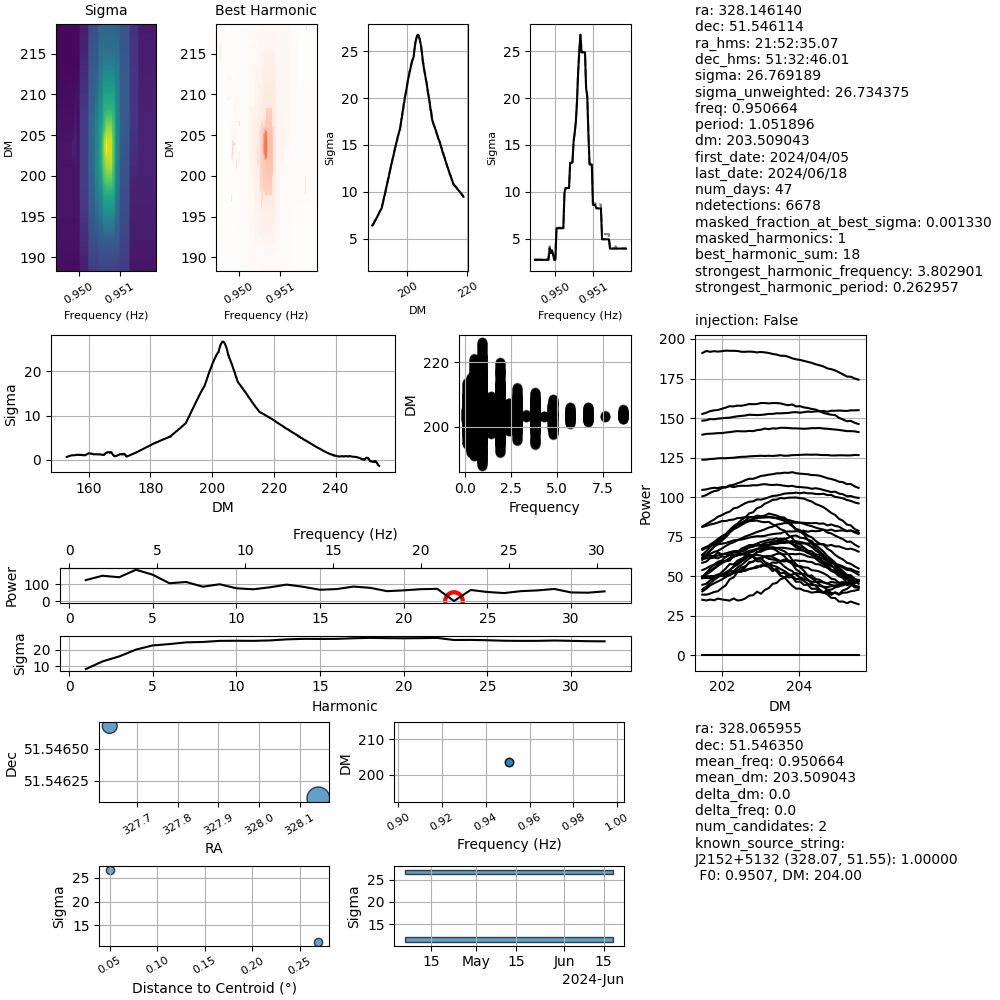}
    \includegraphics[width=0.49\textwidth]{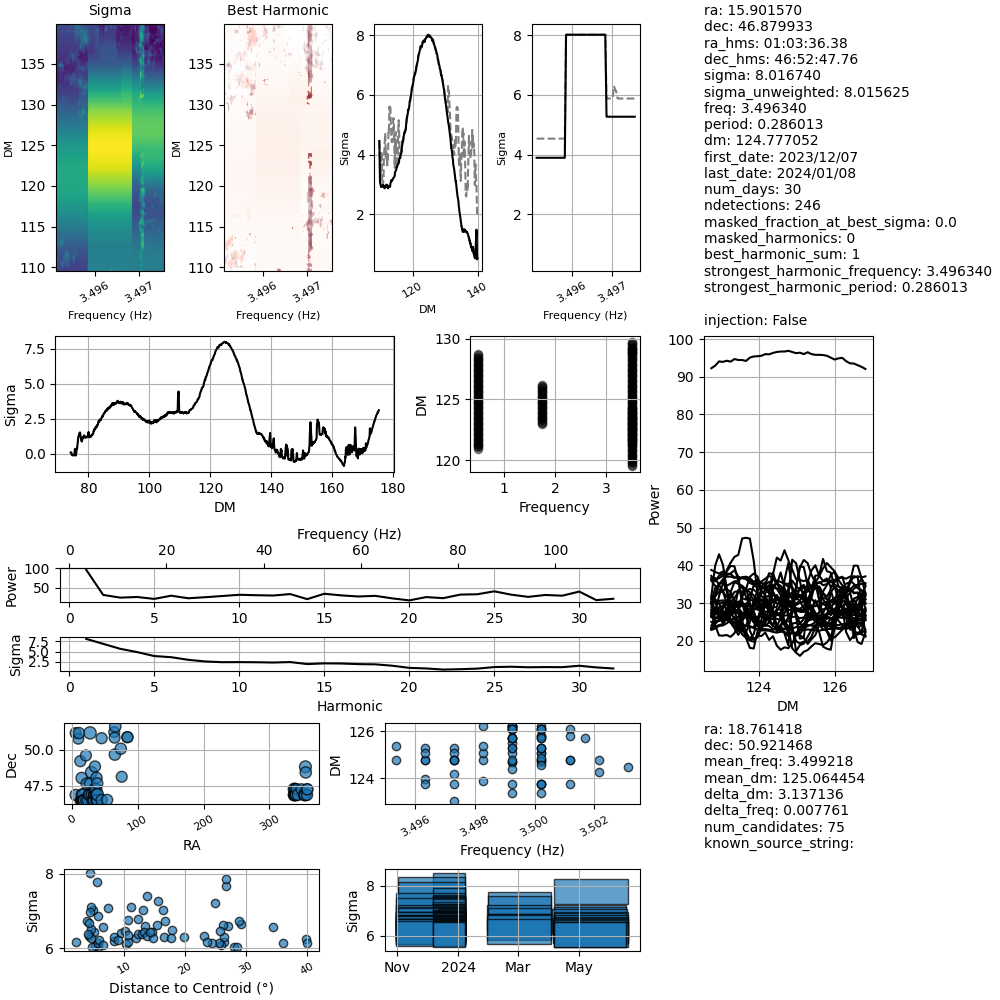}
    \caption{Example of a pulsar (left), and RFI candidate (right). The plots are the result of the multi-pointing clustering process. The upper parts of the plots the show the data of strongest clustered detections and the lower part of the plots show information derived from the multi-poining clustering process. A full description of all fields included in these candidate plots is provided in Section \ref{sec:appendix_cand_plot}. While RFI signals come in many shapes and forms, often they can be discerned from pulsar candidates using the diagnostic plots. This RFI candidate for example shows a spatially broad distribution without a clear central point and the power is concentrated in one harmonic. Often RFI candidates are also very broad or show multiple clear peaks in the DM plane. The pulsar was detected only in two nearby pointings.}
    \label{fig:candidate-examples}
\end{figure*}

The detected signals of pulsars and RFI show a fundamentally different signature across the recorded pointings which is already used in the dynamic birdie filter outlined in Section \ref{sec:powerspectra}. 
While pulsars show up in either only one pointing or a cluster of adjacent pointings with a clear central point, RFI candidates will be spread over a large area of the observed sky with a mostly flat signal strength. We use this difference to perform multi-pointing clustering after the pointings for one day have been processed. This process uses DBSCAN to cluster candidates from multiple pipeline runs by comparing the frequency, DM, right ascension and declination. This creates ``multi-pointing candidates'' which significantly reduce the number of candidates and allows us to further filter them based on their spatial distribution. Examples of multi-pointing candidate plots are shown in Figure \ref{fig:candidate-examples}. The multi-pointing candidates are saved on disk in individual files and a csv file containing a summary of all multi-pointing candidates per-day is created.

The created multi-pointing candidates are matched to known sources by computing the likelihood ratio (i.e. the Bayes factor) of association with
neighboring known sources, based on the values and uncertainties of the candidate and catalog sources.  
This is the same matching scheme used in CHIME/FRB (see \citealt{chimefrb18}), except with an additional parameter of the spin frequency. 
After known source identification the multi-pointing candidates are saved in individual files.

The multi-pointing process is essential in reducing our number of candidates. In a run of our multi-pointing pipeline in December 2024 on the candidates of a search in 10288 power spectra stacks we reduced the total number of candidates from 847,516 to 126,877.

\section{Candidate Confirmation and Follow Up}
\label{sec:followup}

In this Section, we describe our process for candidate confirmation, and follow-up of confirmed candidates.

\subsection{Sifting for Promising Candidates}
\label{sec:foldfilter}
The best way to confirm a candidate is to see a convincing folded pulse profile.  However, due to the nature of our survey, the computational cost is too high to fold every candidate; by the time single-day candidates are formed, the beamformed data have already been removed from memory, so folding requires an additional step of data reading and beamforming per-pointing.  To fold every candidate would be roughly a $1.5\times$ increased compute load.

Instead, we wish to only fold on candidates which are likely to be pulsars. Beginning with a day's worth of multi-pointing candidates, we perform a set of heuristic cuts. 
First, we cut on the significance, DM, and frequency, restricting the candidates to $S/N > 7$, $\rm{DM} > 2\dmu$, $0.01\,{\rm Hz} < f < 100\,{\rm Hz}$, and position spread $\sigma_{\rm position} < 5^{\circ}$ in the multi-pointing candidate.
Candidates matched to known sources are excluded.  
Bright pulsars result in many candidates tightly spaced in DM, and spread among many harmonics in $f$. Not all of these are matched with the known source sifter, so we add an additional cut; all candidates within $1^\circ$, $\sigma_{DM} / \rm{DM} < 0.1$ of a known source are excluded.  

Additionally, unflagged periodic RFI sources often show up at a tight window in $f$ (i.e. compared to the full search space of spin frequency, as they can be broadened due to frequency modulation and barycentreing in the stacks), spread across a large DM range.  We perform an additional cut for these candidates. We create a histogram $N(f)$ of candidates in logarithmic bins of $1\%$ in $f$; for bins with $\geq 2$ candidates with $\sigma_{DM}/\rm{DM} > 0.1$, the candidates are excluded.  Finally, RFI coming from specific sources (electronics, planes, satellites) can cluster in time, and thus RA, which is a proxy for time.  After all other filters, we take only the highest $\sigma$ candidate per pointing.  After the above steps, the characteristic number of daily candidates for 4 beamrows is reduced from $\sim 100000$ to $\sim 200$.

To better tackle this problem as we scale up the survey, we are developing a machine learning classifier, trained on many real candidates, RFI, and injected signals.  This will be discussed in future work.

\subsection{Folding and Multiday $f-\dot{f}$ Search}
\label{sec:folding}
\begin{figure}
    \centering
    \includegraphics[width=1.0\columnwidth]{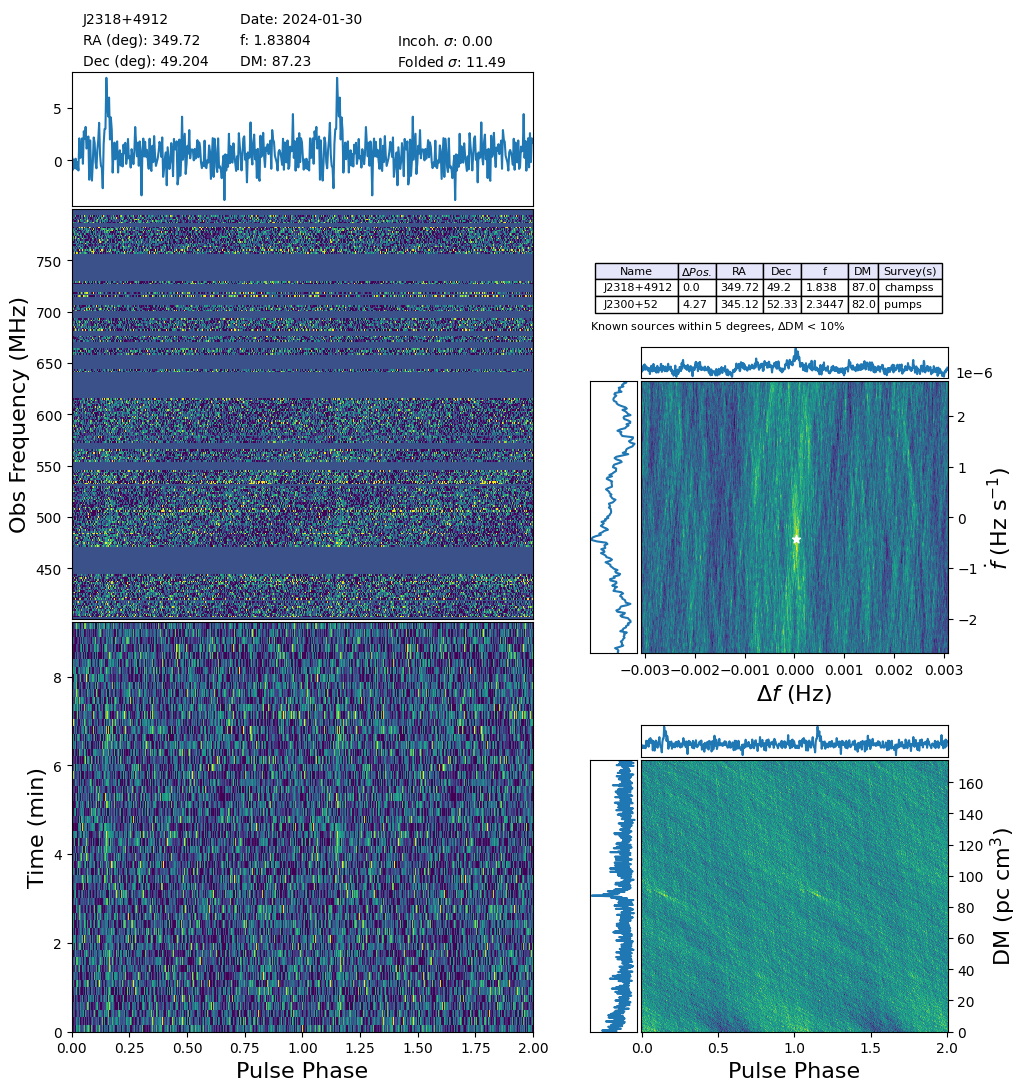}
    \caption{Example of a candidate plot for a single-day fold, on newly discovered pulsar PSR~J2319+4919. The side panels are akin to \texttt{PRESTO} plots, showing the $\chi^2$ as a function of $f$ and $\dot{f}$ (\textit{top}), and $I(\phi, {\rm DM})$ (\textit{bottom}).
    The top contains information of the candidate leading to the fold, and nearby pulsars which might be associated and have passed through the source matching described in Section \ref{sec:multipointing}.
    }
    \label{fig:FoldSingle}
\end{figure}

When a promising candidate is identified, the data corresponding the RA and Dec of the pointing are folded at the candidate $f$ and DM.  For candidates from individual days, the folded candidates are inspected visually, similar to how previous pulsar surveys have operated (see Figure \ref{fig:FoldSingle}).

For candidates only detected in the power spectrum stack, the pulsar is likely too faint to detect in a single day; all of the existing data on disk, and each subsequent day, are folded with the same candidate ephemeris, and summed in time to form $I(T_i, \phi)$, where $\phi$ is the pulse phase and $T_i$ is the central time of the $i$th observation.
The candidate $f$ derived from minutes-length transits is too imprecise to phase align pulses from day to day.  
Moreover, a pulsar spin-down, and uncertain pulsar position result in a time-variable spin frequency.  
Over 1 month, for an isolated pulsar these effects can all be approximated sufficiently well with with a spin frequency derivative $\dot{f}$ which is allowed to be positive or negative, and the phase is approximated as
\begin{equation}
    \phi(t) = f t_1 \approx f_0 t_1 + \frac{1}{2} \dot{f} t_1^{2},
\end{equation}
where $t_1 \equiv t - t_{\rm ref}$, $f_0$ is the spin frequency at the reference time $t_{\rm ref}$, set to the central observation.  
Binary pulsars add an additional complication, see Section \ref{sec:binaries}.

\begin{figure}
    \centering
    \includegraphics[width=0.95\columnwidth, trim={0cm 0 16.8cm 3.3cm}, clip]{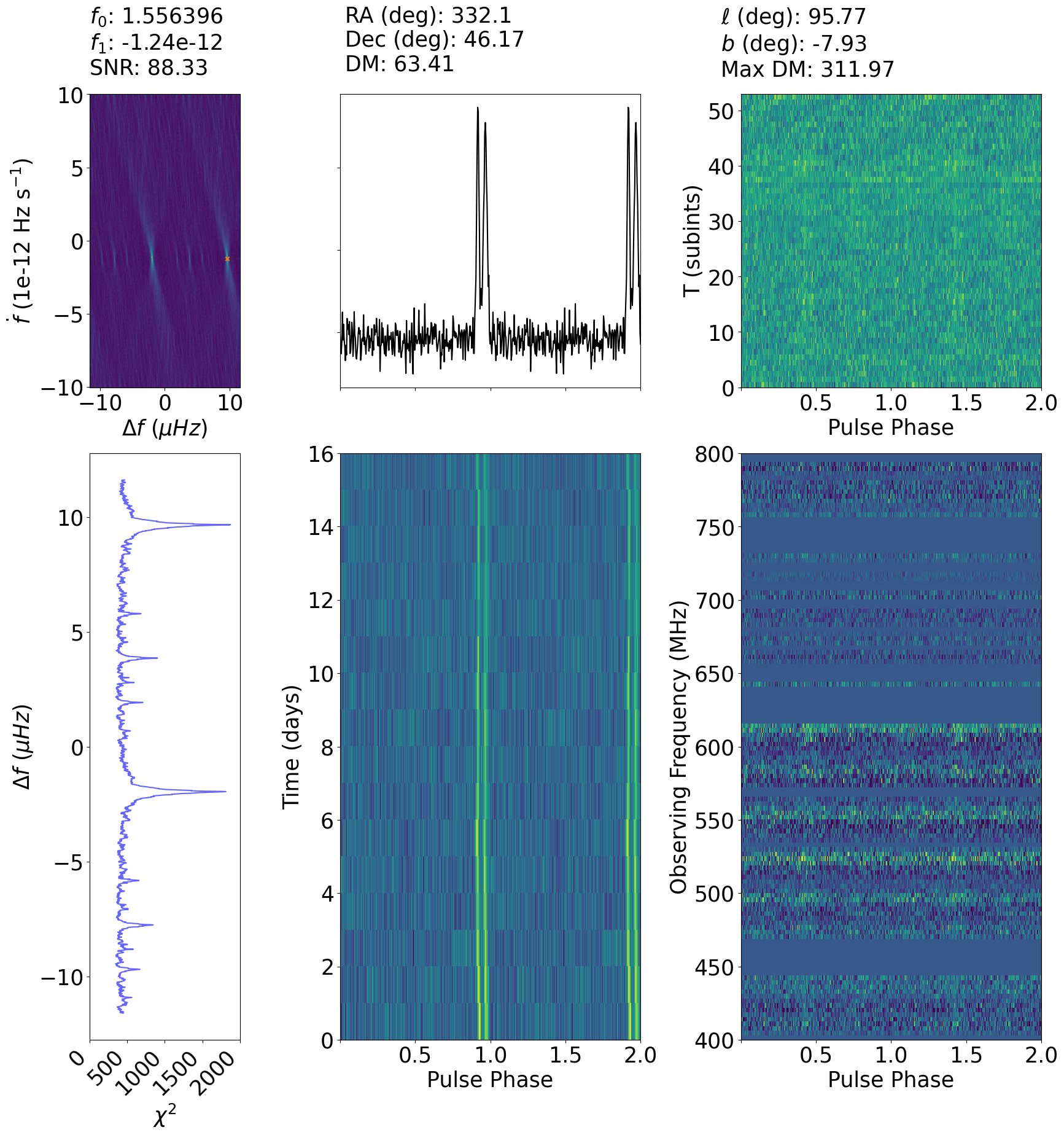}
    \caption{An example of our multiday search script for candidate confirmation, which searches a grid of $f$, $\dot{f}$ values as described in Section \ref{sec:folding}. All of our newly discovered pulsars were found with high significance from this algorithm. Shown above is known pulsar PSR~J2208+4610 \protect{\citep{dong+23}} which is in a 412.5\,day orbit, demonstrating how the search can phase align systems in long orbits. Typical isolated pulsars are found with smaller $\dot{f}$ owing to spin-down and position uncertainties.}
    \label{fig:multiday}
\end{figure}

When the requisite number of days of data have been folded (a tunable parameter, set to 10 days thus far), a grid of $f$, $\dot{f}$ values are searched, resulting in $\chi^2(f, \dot{f})$. 
The spacing in both $f$ and $\dot{f}$ are chosen to be uniform in phase, with $\Delta f$, $\Delta \dot{f}$, corresponding to a one bin phase shift between the first and last observations.  
We set the maximum search frequency of $\Delta f \approx 11.6\,\mu\rm{Hz}$, corresponding to $1 / T_{\rm sid}$.  
This is guaranteed to contain a value of $f$ which can align the pulses in phase;  due to the transit nature of CHIME observing at the same sidereal time, there will be a family of aliased solutions separated by integer number of $N$ pulses per sidereal day.  
We search to a maximum frequency derivative of  $\dot{f} = 10^{-12}\,\rm{Hz}\,\rm{s}^{-1}$, which is sufficiently large to contain most pulsars, and captures any intra-beam positional uncertainty.  
If no signal is seen in this grid, then $\dot{f}_{\rm max}$ is increased by a factor of 10 and the search re-run, up to $\dot{f}_{\rm max} = 10^{-10}\,\rm{Hz}\, \rm{s}^{-1}$.  
An example of the multiday search is shown in Figure \ref{fig:multiday}.

All of the above described confirmation uses the CHAMPSS data themselves (i.e. from the CHIME/FRB datastream).  More rapidly rotating sources will benefit from CHIME/Pulsar follow-up (see discussion in Section \ref{sec:sensitivity}).

\subsection{Timing}
\label{sec:timing}

\begin{figure*}
    \centering
    \includegraphics[width=\linewidth]{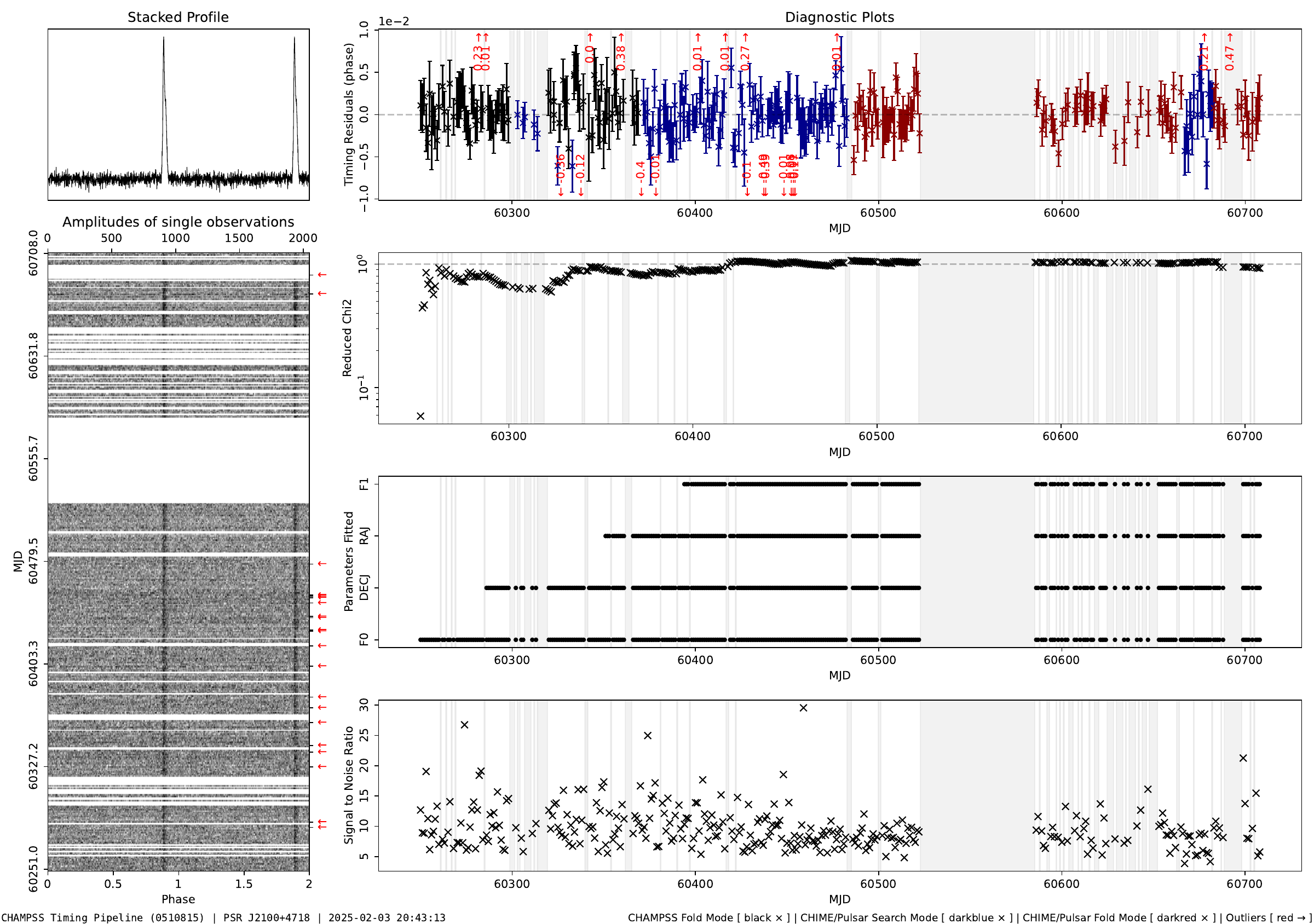}
    \caption{An example output from the timing pipeline described in Section \ref{sec:timing} with PSR~J2100+4711. \textit{Left:} Pulse intensity (greyscale) as a function of phase and date of pulse profiles after the timing model has been applied (bottom), and their stacked profile by summing over time (top). The noise level depends on whether CHAMPS or CHIME/Pulsar data was used. The panels on the right (from top to bottom) show the timing residuals, reduced $\chi^2$, parameters being fitted for, and the S/N per day.  For the residuals, black points correspond to CHAMPSS, blue points to CHIME/Pulsar filterbank (search-mode) data, and red points to CHIME/Pulsar fold-mode data. Red arrows indicate outlier TOAs, whose residual is given by the number next to it. Omitted for space, the viewer shows a panel with the full ephemeris and uncertainties on free parameters.}
    \label{fig:timingpanorama}
\end{figure*}

When a new pulsar candidate is confirmed, it is observed by CHIME/Pulsar using the ephemeris derived from the search pipeline. Since CHIME/Pulsar (mainly due to its tracking beam) provides a higher sensitivity than CHAMPSS, most timing TOAs are obtained from the CHIME/Pulsar fold-mode backend. However, as CHIME/Pulsar uses a probabilistic scheduling system \citep{chimepulsar21}, daily observations are not guaranteed. Thus, CHAMPSS fold-mode data are used to start the timing model (i.e. before the pulsar is queued into CHIME/Pulsar schedule), and to fill gaps between CHIME/Pulsar TOAs.

We developed a pulsar timing pipeline to time pulsars on a daily basis. To start the pipeline, we create an initial standard template for each pulsar using \texttt{paas}; this template is later replaced by an averaged profile or a \texttt{fitburst}\footnote{\url{https://github.com/CHIMEFRB/fitburst}} \citep{fonsecaModelingMorphologyFast2024} modeled template. In the pipeline, data are first masked for bad channels and cleaned using the \texttt{clfd} RFI removal algorithms\footnote{\url{https://github.com/v-morello/clfd}} \citep{mbc+19}. TOAs are obtained using \texttt{pat} after summing the profile in frequency, time, and polarization using \texttt{pam} (\texttt{paas}, \texttt{pat}, and \texttt{pam} are all part of \texttt{PSRCHIVE}\footnote{\url{https://psrchive.sourceforge.net/}}, \citealt{hotan+04}). Lastly, timing solutions are fitted using the \texttt{PINT}\footnote{\url{https://github.com/nanograv/PINT}} \citep{luo+21} least-squares fitter, with TT(BIPM2021) \citep{pet+10} as the reference clock standard and DE421 \citep{fwb+09} as the solar system ephemeris (automatically been used by \texttt{PINT}). For this paper, the timing solutions were re-fitted using \texttt{PINT} MCMC fitter to account for parameter degeneracies and presented in Table \ref{tab:resultstable}.

This timing procedure also automatically adds parameters to the timing model by running F-tests, similar to the `Algorithmic Pulsar Timing' scheme proposed by \citet{phillips+22}. Once the degeneracy in a timing model is sufficiently constrained (typically after $\sim$1 year of timing, as inferred from the MCMC posterior distributions), the initial ephemeris used in the CHIME/Pulsar fold-mode backend is replaced by the newly fitted ephemeris. The updated ephemeris provides an improved timing position and an unaliased period solution (will be further introduced in Section \ref{sec:resolving_aliased_solution}), resulting in a higher signal-to-noise ratio for fold-mode observations. An example diagnostic plot from our automatic timing pipeline is shown in Figure \ref{fig:timingpanorama}.

\vspace{1cm}
\subsection{Resolving Aliased Solutions} \label{sec:resolving_aliased_solution}

As mentioned in Section \ref{sec:folding}, due to the transit nature of CHIME, the multi-day phase connecting algorithm allows for a family of solutions with pulse periods $\pm N$ pulses/day, for any integer $N$.  A period off by $N$ rotations/day will drift across a transit by $\Delta \phi = N T_{\rm transit}/T_{\rm sid}$.  For example, a solution off by 1 pulse/day, and a 10\,minute transit has a phase drift of $\Delta \phi \approx 0.007$ across each transit.  This is a small effect, but measurable given high S/N; see, e.g., the phase errors on individual TOAs in Fig \ref{fig:timingpanorama}.

\begin{figure}
    \centering
    \includegraphics[width=1.0\columnwidth]{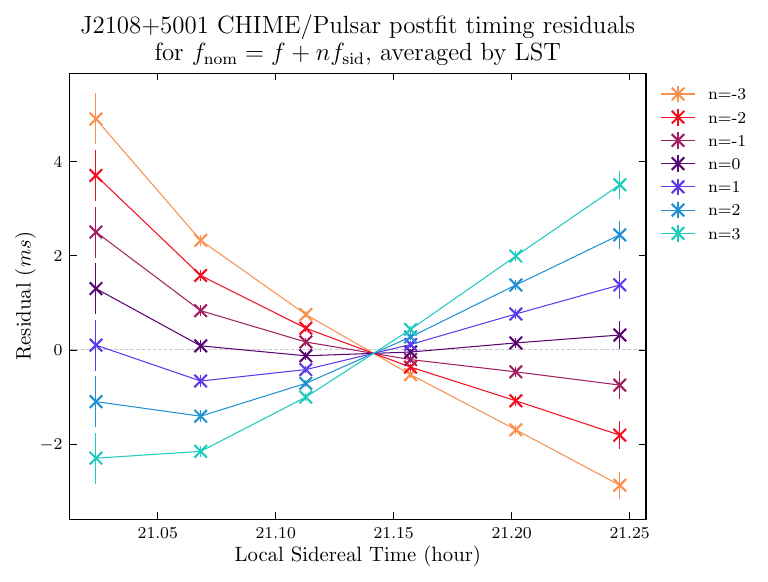}
    \caption{LST-averaged residuals for aliased timing solutions for J2108+5001. For each line, the frequency in the timing solution was adjusted by $nf_\mathrm{sid}$, TOAs were then refitted, and residuals at similar local sidereal times were averaged together. A slope indicates residuals shift in phase over the course of an observation and, therefore, that the frequency is an alias of the true value.}
    \label{fig:alias-lst-plot}
\end{figure}

For pulsars which are bright enough to be detected in a single day, multiple TOAs are made for each observation. This allows for the frequency to be fitted based on the drift across a transit. If necessary, JUMPS are added between each observation to fit for the drift across all transits alone, without complications due to other factors such as position. This frequency can then be used to derive a new timing solution. These TOAs can also be used as a diagnostic to check whether an existing timing solution contains an aliased frequency. To do this, the frequency is adjusted by a multiple of the sidereal frequency $f_\mathrm{sid} = 1/T_{\rm sid}$ and the timing parameters refitted. By plotting the residuals as a function of LST a clear trend can be seen for the aliased solutions, especially after averaging residuals at similar LSTs. An example of this is shown in Figure \ref{fig:alias-lst-plot}.  For pulsars which are not bright enough to be seen in individual days, the above process can still be done. Each profile traverses the same sidereal time; after first timing with a single time, frequency averaged TOA per day, profiles can be stacked to form $I(t_{\rm LST})$, and TOAs can be extracted from this stacked profile to form TOAs as a function of LST.

\subsection{Binaries}
\label{sec:binaries}

We may often find that a binary pulsar is bright enough to be seen in the power spectrum stack, but not bright enough to show up as candidates in individual days.  In these cases, as a candidate $f$ value will be close to the average over the (unknown) orbit, and will lead to phase wrapping even in single observations.  Moreover, tight high-velocity orbits will lead to phase evolution in single days.

We describe several ways we enable sensitivity to binaries.

\paragraph{$P_{b} \gtrsim 100$\,days}
This is the easiest case for our survey, where the orbital phase is well captured by a frequency derivative over the course of the 10 day phase coherent search.  The $\dot{f}$ term can then capture the binary motion, allowing us to phase connect the pulsar. For an example, see left panel of Figure \ref{fig:multiday}.  Using the Keplerian parameters of all known pulsars in the ATNF Catalogue \citep{manchester+01} with binary periods greater than 100\,days, and rotation frequency less than 100\,Hz,
the maximum and mean induced acceleration $\dot{f}$ throughout the orbit are less than $10^{-8}\,\rm{Hz}\,\rm{s}^{-1}$ and $10^{-9}\,\rm{Hz}\,\rm{s}^{-1}$, respectively.

\paragraph{$P_{b} \lesssim 100$\,days}
This case is more difficult to detect in our survey.  On individual days, we search a range of $\Delta f$, $\dot{f}$ values, 
setting the maxima to
\begin{equation}
    \Delta f_{\rm max} = f v_{\rm max}/c
\end{equation}
\begin{equation}
    \dot{f}_{\rm max} = \frac{2\pi}{P_{b, {\rm min}}} \Delta f_{\rm max}.
\end{equation}
Since we fold comparatively few candidates, we can search a large range in $\Delta f$, $\dot{f}$ without much additional computational pressure.  We set the values using $v_{\rm max} = 1000\,$km/s, $P_{\rm b, min} = 2\,$hours.
With this scheme, even tight, high-velocity orbits can be detected if the pulsar is sufficiently bright to be seen in individual days.  A starting estimate for the orbital parameters can be obtained by fitting $f(t)$, although this can also lead to aliased solutions due to observing at the same sidereal time each day.  A significant measure of $\dot{f}$, or an observation offset from CHIME's observing window can immediately break the aliasing degeneracy.  

To phase connect binary pulsars, TOAs in single days are produced as described in Section \ref{sec:timing}.  Then the incoherent fit for the binary orbital parameters is included in the timing model, and inspected for correlated structure in the TOAs and to see if the solution needs phase jumps applied.  The requirement for the pulsar to be visible in individual days, and the benefit of being able to measure $\dot{f}$ from orbital acceleration, makes CHIME/Pulsar the preferred instrument for these techniques.  

\begin{figure}
    \centering
    \includegraphics[width=1.0\columnwidth, trim={0cm 4.5cm 0cm 0cm}, clip]{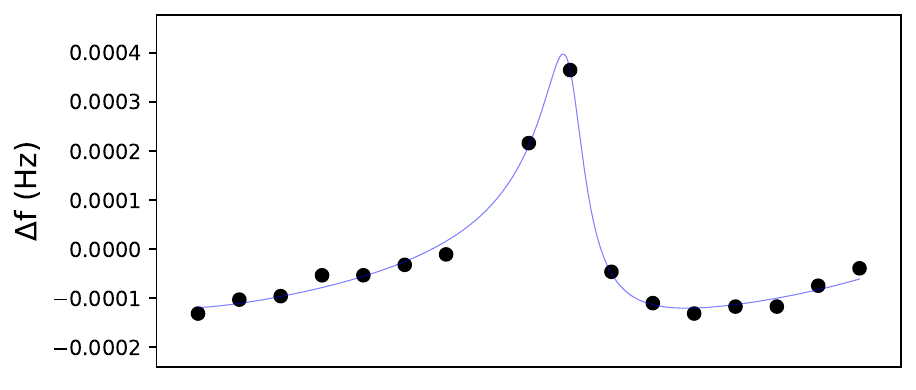} \\
    \hspace{1.5mm}\includegraphics[width=0.98\columnwidth]{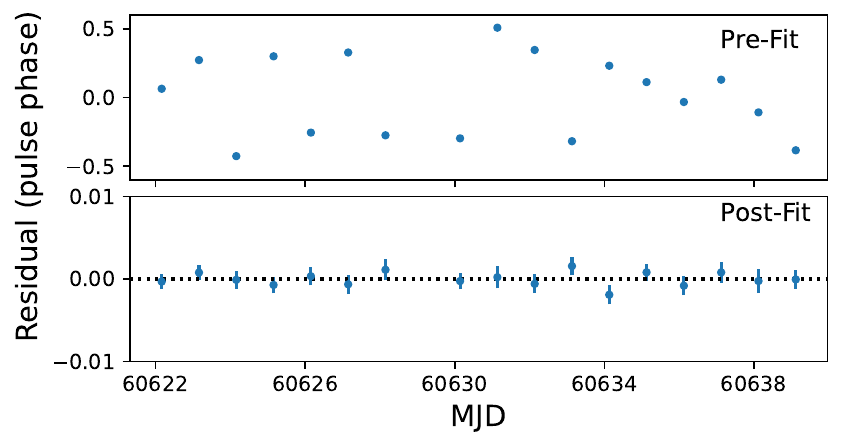} \\

    \caption{(\textit{Top:}) Varying period found from the single-day candidate folds 
    of PSR~B2303+46, in an eccentric ($e = 0.658$) 12.34\,day orbit.  The pulsar was treated as an unknown candidate, folded with a constant period from the peak of $f$ in the stack, and folded daily using the standard $f-\dot{f}$ grid search for candidates. The fit orbital solution was close enough that the pulsar could be phase aligned using \texttt{pintk}, assigning jumps to phase wrapping TOAs, with pre- and post-fit residuals shown in the bottom two panels. Note the vastly different scales; The post-fit residuals contained within $<0.5\%$, while pre-fit are scattered across the entire pulse phase.}
    \label{fig:orbitdF}
\end{figure}

As proof of concept, we tested our algorithm on PSR~B2303+46, which was in the declination range of our commissioning survey.  This pulsar has a spin frequency of $f \approx 0.938$\,Hz, and is in a $12.34$-day orbit with eccentricity $e=0.658$. The pulsar was folded blindly on its candidate peak $f$, DM, from the power spectrum stack; the daily best-fit $f$, least-squares orbital fit, and final phase-connected timing solution are shown in Figure \ref{fig:orbitdF}.

\paragraph{Caveat on detecting binary pulsars in power spectra stacks}  The aforementioned strategies both assume that the source is sufficiently bright to be detected either in the single-day power spectra, or in the stacks.  However, the binary motion can have the effect of smearing the signal by multiple power spectrum bins, decreasing S/N.  The signal will be smeared by
\begin{equation}
\Delta f = \frac{v_r}{c} f_0,
\end{equation}
scaling both with the radial velocity of the orbit, and the spin frequency.  This effect will thus be most deleterious for rapidly rotating pulsars in tight binaries.  

In later iterations of the survey, we will include some strategy to search for orbitally modulated signals in the power spectra stacks.  For an orbit with $P_b \ll T_{\rm stack}$, with sufficiently many observations the power distribution in the stack will reflect the integral of $v_r/c$ throughout the orbit.  These could be searched for using additional binning, or matched filters; we leave this to future work.

\section{Operations}
\subsection{Workflow}
\label{sec:workflow}

Originally developed for the CHIME/FRB team, Workflow is an in-house framework designed to manage the lifecycle of processes, including execution, pipelining, result saving, and querying. It is agnostic to underlying parameters such as hardware, software, compute environment, or processing constraints. Workflow provides a command-line interface (CLI) and Python modules that enable users to queue tasks (``Work'') into collections (``Buckets'') in a MongoDB database. These tasks can then be processed through the CLI or Python, with the results automatically stored in a separate MongoDB collection (``Results'').

An additional feature, Pipelines, allows users to define sequential or concurrent tasks in YAML files, specifying schedules and resource requirements such as CPU and RAM. Pipelines integrate with Docker Swarm\footnote{\url{https://docs.docker.com/engine/swarm/}}, a multi-node orchestration tool for managing containerized workloads, to automate the scheduling and execution of tasks based on available resources. This ensures efficient parallel processing across nodes meant to handle the entire workflow$-$task submission, container scheduling, execution, and result storage.

The Workflow Web interface provides
real-time status monitoring of Buckets, Pipelines, and Results, as well as advanced querying and visualization of outputs, including plots linked from stored file paths.

We utilize Docker Swarm's replicated mode by creating a Docker Service for each pointing, where each service corresponds to a single Docker Container (referred to as a ``replica'' within Docker services, which typically have many replicated containers per its service). Using one replica per service is deliberate, due to each pointing process requiring a unique CPU and memory reservation that cannot be replicated across a single service.

To coordinate operations, a manager service determines the available daily processes based on the specified raw data folder, and uses the Docker SDK for Python to launch single-pointing jobs as services. A cleaner service monitors for completed tasks (successful or failed), outputs logs to a designated folder, and removes the completed services from the Docker Swarm state.  Currently, this is designed to operate in quasi-real time, determining and running all processes for a given day and stepping through days sequentially (with the ability to process any day for which there are data on disk)

Each service is assigned CPU and memory reservations according to an empirically derived formula based on the maximum DM per pointing. Docker Swarm manages job scheduling by comparing the cumulative memory reservations against the total available memory of the node, determining whether a job should be ``pending'' or ``running''. The allocation of jobs across nodes follows a round-robin strategy based on available resources. To enforce task ordering, necessary for the RFI algorithm, the manager service waits if any 1 job is in a ``pending'' state.

Given that each CHAMPSS compute node has a CPU with 128 logical cores 
paired with 256\,GB of RAM (described in \ref{sec:computing}), the number of threads per pointing process is set to the reserved memory (in GB) divided by three, as CPU usage scales proportionally with memory usage. This allocation prevents CPU over-scheduling, ensuring optimal performance for other system tasks that need threads (hence why the reserved memory is not strictly divided by two) and processing tasks (that benefit from having fully-available threads). Each container starts by using the Workflow CLI to poll for jobs in its Workflow Bucket and completes once the pointing process finishes. The container is booted up from an image which, on the first job for a node, is pulled from a local Docker Image registry. The registry is updated using GitHub Actions upon any modification to the main GitHub branch of our codebase.

This setup also applies to other CHAMPSS tasks, including the multi-pointing candidate grouping (Section \ref{sec:multipointing}), and folding (Section \ref{sec:followup}), which are run daily after all individual pointings have been processed.
In future iterations, a single YAML file will define the entire CHAMPSS pipeline, specifying Python modules and functions, their execution order, and jobs counts per step. This YAML configuration will be deployed to an on-site Workflow Pipelines server running within a Docker Service. Workflow will then manage this full process lifecycle automatically for the team. Currently, the team employs Workflow Buckets, Results, and the Web interface only, while integration of our custom Docker Swarm scheduling into Pipelines is ongoing.

\subsection{Metadata Database}
\label{sec:database}

Our pipeline uses a local MongoDB database. This database contains the following collections:

\begin{itemize}
    \item Pointings in pointing map
    \item Available pipeline processes
    \item Processed observations
    \item Existing power spectra stacks
    \item Known sources
    \item Sources which are followed up
\end{itemize}

These collections are accessed and updated during the various processing steps of the pipeline. This database is separate from the the Workflow database (see Section \ref{sec:workflow}). 

The known sources are first populated using the ATNF database, and updated by querying the Pulsar Survey Scraper \citep{pulsarscraper}.

\vspace{0.5cm}
\subsection{Candidate Viewer}

\begin{figure}
    \centering
    \includegraphics[width=1.0\columnwidth, clip]{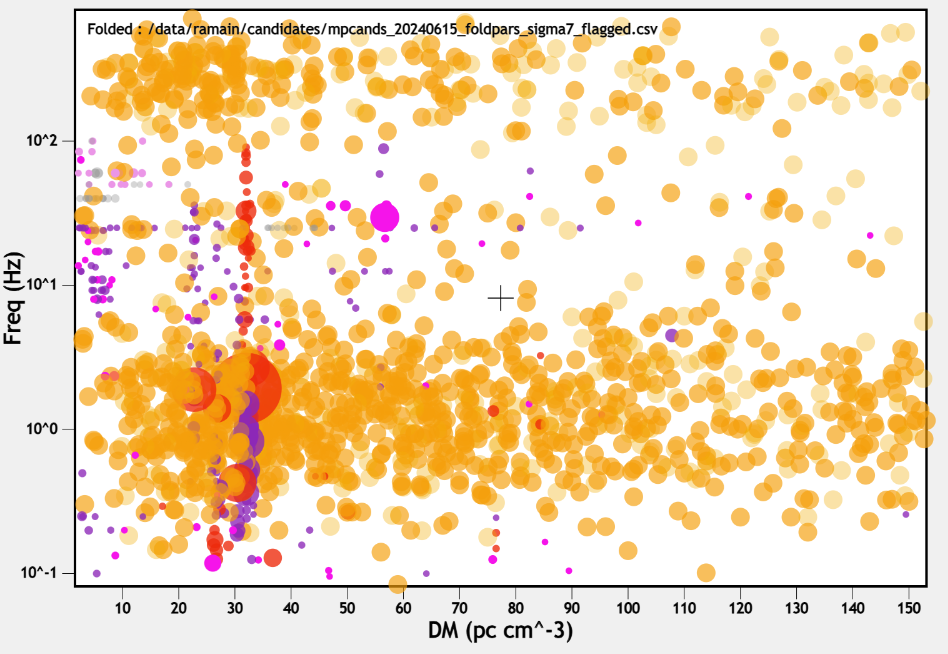}
    \caption{Example data visualization for a given day's data product. Orange points represent known pulsars from the PSRCAT database, purple and pink points are multi-pointing candidates from the pipeline, red points are candidates that were associated with a known source. The radius of each circle scales with the $\sigma$ of that candidate.}
    \label{fig:candidateViewerDemo}
\end{figure}

In order to visualize our daily and monthly data products, a Candidate Viewer (shown in Figure \ref{fig:candidateViewerDemo}) was developed to map the distribution of multi-pointing candidates across the sky. Constructed using a browser-based framework called \href{https://p5js.org/}{p5.js}, this tool can load summary csv files to represent these candidates as points in the user-defined phase space (RA/Dec, Frequency/DM, etc.). The customizable viewing port then allows users to visually compare candidates to each other and to known pulsars with similar values. The data can also be filtered and sorted dynamically by any parameter.

If the user finds an interesting candidate, they can select it to get a more detailed overview of its properties and associated plots, with the option to give the candidate a label shared with the rest of the team. Furthermore, since the tool can interface with Workflow, users can queue a folding job on the candidate to the computing cluster, making it easy to find and follow up on promising sources manually.

While the candidate viewer was developed specifically for CHAMPSS, it can be broadly applied to any table with coordinates, numerical values, and optional metadata.  The viewer is shown in detail, along with a code release, in an accompanying research note (Laurent Tarabout 2025 - reference added when this paper is published).

\section{Commissioning Surveys}
To test the system, we have performed multiple commissioning surveys including only a subset of the sky.

The first survey was performed in December 2021/January 2022. Data were recorded for 34 days in beamrows 125-132 and RA range 313.7-322.3 degrees. The data were shipped off-site for further analysis as, at that point, the pipeline was not fast enough to run in real time and the on-site CHAMPSS processing nodes did not exist. Unfortunately, these were affected by severe RFI so additional data were taken for 11 days in June 2022.
The analysis of these data resulted in the discovery of J2108+5001.

Our CHAMPSS commissioning cluster was installed at the CHIME site in 2022 (see Section \ref{sec:computing}).
With this cluster, quasi-realtime processing is possible for a small fraction of the CHIME sky.  We performed a realtime commissioning survey which continuously recorded data for over a month from each beamrows 120--123, 124--127, 128--133, sequentially, lasting from October 2023 to June 2024 (56 beams total).  These data were automatically searched daily, as well as the $\sim$ 1--2 month incoherent stack.  A footprint of this commissioning survey is shown in Figure \ref{fig:pointing-map}.

\subsection{Results}

\begin{figure}
    \centering
    \includegraphics[width=1.0\columnwidth]{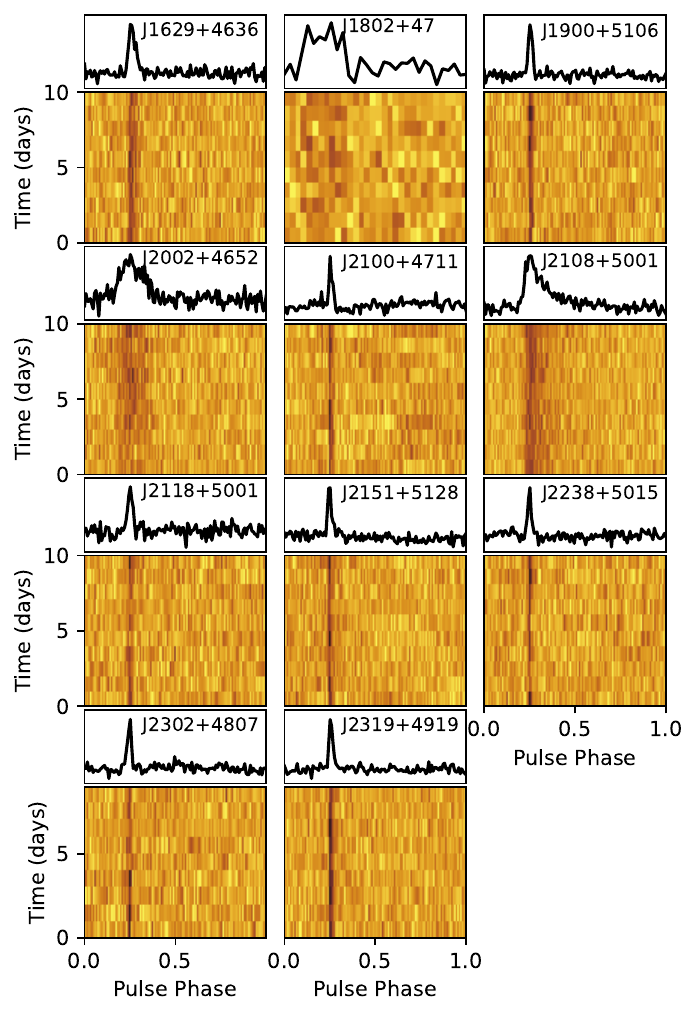}
    \caption{Panorama of newly discovered pulsars. The pulsars have been averaged in frequency across the band, binned to 128 phase bins, and averaged over 10 days.}
    \label{fig:panorama}
\end{figure}

The commissioning surveys have resulted in the discoveries of 11 new pulsars.  The newly discovered pulsars are all isolated, spanning a period range of $0.24 - 1.46\,$s.  Profiles of our newly discovered pulsars over 10 days are shown in Figure \ref{fig:panorama}, and their basic properties are listed in Table \ref{tab:resultstable}. Of the 11 pulsars, 6 have sufficient follow-up observations to derive their timing solutions at the time of writing. Their parameters fitted by our timing pipeline are also listed in the Table \ref{tab:resultstable}, and their timing residuals are shown in Figure \ref{fig:timing_residuals} in the appendix. New pulsars and updated ephemerides will be updated to our website\footnote{\url{https://chime-sps.github.io/pulsars_webpage/pulsars/}}.

To estimate the flux densities of our newly discovered pulsars, we start with CHIME/Pulsar pulse profiles, generated as described in Section \ref{sec:timing}. The off-pulse mean ($\bar{p}$) of these uncalibrated profiles corresponds to the system equivalent flux density (SEFD) of our telescope. We can calibrate our profile, $I(\phi)$ by subtracting then dividing by $\bar{p}$, then multiplying by ${SEFD}$ to convert it to units of Jy:
\begin{equation}
    I(\phi)_{cal} = SEFD\left( \frac{I(\phi) - \bar{p}} {{\bar{p}}}\right)
\end{equation}
The mean flux density is then simply the area under $I(\phi)_{cal}$ divided by the number of phase bins. To use this method, we need the SEFD for each pulsar, for which we use calibrator radio sources monitored by CHIME/Pulsar, as outlined in Section 3.5 of \citet{Dong_2024}. Measuring the calibrator sources when they are on sky and then off sky and comparing their catalogued flux densities, we can get the system temperature, gain, and thus the SEFD, for the pointings corresponding to each pulsar. We apply the frequency-averaged SEFD values to the uncalibrated profiles and calculate a mean flux density $S_{600}$ for each pulsar, listed in Table \ref{tab:resultstable}. The weakest pulsar we detect is J2302+48, with a flux density of 0.24 mJy.

\begin{table*}[ht]
    \centering
    \label{tab:resultstable}
    \setlength{\tabcolsep}{3pt}
\begin{tabular}{|c|cccc|cc|cc|}
\hline
\multirow{2}{*}{PSR} & \multicolumn{4}{c|}{Timing Solution} & \multicolumn{2}{c|}{Derived Pars} & \multicolumn{2}{c|}{Pointing Position} \\ \cline{2-9} 
 & R.A. & Dec & $P$ & $\dot P$ & DM & $S_{600}$ & R.A. & Dec \\
 & \scalebox{0.9}[0.9]{(hh:mm:ss)} & \scalebox{0.9}[0.9]{(hh:mm:ss)} & \scalebox{0.9}[0.9]{(s)} & \scalebox{0.9}[0.9]{(s s$^{\rm -1}$)} & \scalebox{0.9}[0.9]{(pc cm$\rm^{-3}$)} & \scalebox{0.9}[0.9]{(mJy)} & \scalebox{0.9}[0.9]{(hh:mm:ss)} & \scalebox{0.9}[0.9]{(hh:mm:ss)} \\ \hline
J1629+4636 & 16:29:$52.904(3)$ & +46:36:$51.75(2)$ & $0.314056114549(5)$ & $8.51(2)\times 10^{-17}$ & 34.8 & 0.51 & 16:29:53 & 46:36:53 \\
J1802+47 & 18:02:$14.4(5)$ & +47:16:$00(^{+10}_{-50})$ & $0.346624379(4)$ & $1.1(3)\times 10^{-15}$ & 30.1 & 0.31 & 18:01:27 & 47:22:50 \\
J1900+5106 & 19:00:$0.92(7)$ & +51:06:$14(1)$ & $0.3377591590(8)$ & $3.2(3)\times 10^{-16}$ & 71.6 & 0.46 & 19:00:19 & 51:13:40 \\
J2002+4652 & 20:02:$06.87(9)$ & +46:52:$43(4)$ & $0.248260284(1)$ & $2.6(7)\times 10^{-16}$ & 141.2 & 0.76 & 20:02:00 & 46:56:20 \\
J2100+4711 & 21:00:$13.393(^{+7}_{-8})$ & +47:11:$15.4(1)$ & $1.45874256105(7)$ & $4.139(3)\times 10^{-15}$ & 231.1 & 0.31 & 21:00:13 & 47:11:10 \\
J2108+5001 & 21:08:$41.962(3)$ & +50:01:$41.73(5)$ & $0.24446137520(2)$ & $9.1442(4)\times 10^{-15}$ & 482.3 & 0.89 & 21:08:51 & 50:00:30 \\
J2118+5143 & 21:18:$30.26(7)$ & +51:43:$15.4(6)$ & $0.3702230490(5)$ & $3.470(2)\times 10^{-14}$ & 146.3 & 0.61 & 21:18:54 & 51:33:24 \\
J2151+5128 & 21:51:$46.60(5)$ & +51:28:$48.9(2)$ & $1.0519028955(6)$ & $9(3)\times 10^{-17}$ & 203.5 & 0.73 & 21:52:15 & 51:32:46 \\
J2238+5015 & 22:38:$16.49(6)$ & +50:15:$52.4(3)$ & $0.5600971676(5)$ & $4(2)\times 10^{-17}$ & 28.3 & 0.59 & 22:37:58 & 50:22:20 \\
J2302+4807 & 23:02:$10.9(4)$ & +48:07:$26(4)$ & $0.741973791(8)$ & $1.3(3)\times 10^{-15}$ & 72.7 & 0.24 & 23:02:36 & 48:02:33 \\
J2319+4919 & 23:19:$14.842(3)$ & +49:19:$8.72(8)$ & $0.54406513568(2)$ & $1.26(1)\times 10^{-16}$ & 87.2 & 0.81 & 23:19:10 & 49:24:05 \\ \hline
\end{tabular}
    \caption{Parameters of newly discovered pulsars are shown, including the follow-up timing solutions (left), derived parameters (middle), and detection positions (right).
    The timing solution was fitted by \texttt{PINT} using its MCMC fitter, and the upper and lower bounds are given by 16th and 84th percentiles of the posterior distribution (a single value is shown when the distribution is symmetric), respectively. Reduced $\chi^2$ values for these timing solutions are shown in Figure \ref{fig:timing_residuals}. More technical information about timing can be found in section \ref{sec:timing}. Note that EFAC was not applied to rescale the uncertainties during the fit.}
\end{table*}

\subsection{Known Pulsars, and System Performance}
\label{sec:sensitivity}

To better quantify our sensitivity, we use detections of known pulsars that lie within the RA/Dec range of our commissioning survey and have previously known flux densities. Using the radiometer equation we calculate the expected minimum detectable flux density for each known pulsar, 
\begin{equation}
\text{$S_{min}$} = \frac{\text{$S/N_{min}$} (\text{$T_{rec}$} + \text{$T_{sky}$})}{G\sqrt{n_p  
\text{$t_{int}$} \Delta f}}  \sqrt{\frac{W}{P - W}},
\end{equation}
where minimum detection threshold $S/N_{min} = 7$, average gain at zenith $G=1.16 \hspace{0.1cm} \text{K Jy$^{-1}$}$ \citep{good+21}, $n_p=2$ is the number of polarizations, $t_{int}$ is the integration time, $\Delta f=200 \hspace{0.1cm}\text{MHz}$ is the effective bandwidth\footnote{CHIME has a native bandwidth of $400 \hspace{0.1cm}\text{MHz}$. The bad channel and the RFI cleaning in CHAMPSS remove $\sim50\%$ of the bandwidth, leaving $\Delta f = 200 \hspace{0.1cm}\text{MHz}$. }, $W$ is pulse width, and $P$ is spin period. 

\begin{figure}
    \centering
    \includegraphics[width=1.0\columnwidth, clip]{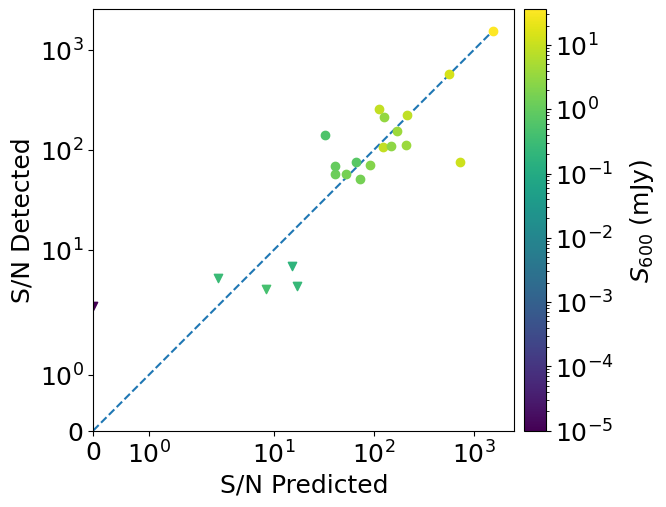}
    \caption{Comparison of expected vs. actual signal-to-noise values for folded known pulsars that lie within the CHAMPSS commissioning survey RA/Dec range. An upper limit is indicated for pulsars that are not detected by CHAMPSS and the dashed line is the line of equality.}
    \label{fig:sensitivity}
\end{figure}

The process is similar to the one outlined in section 3.3.4 of \citet{good+21}. We assume a different receiver temperature of 50\,K based on the average of calibrator sources used in the SEFD calibration method of \citet{Dong_2024}, as the previous value of 30 K has consistently been shown to underestimate flux densities of known pulsars. For the sky temperature we use a global sky model from \texttt{pygdsm} that combines several surveys, as recommended by \citet{price+21}. This global sky model includes the 408\,MHz all-sky map of \citet{haslam}. We use $S_{600}$ values from the Australian Telescope National Facility (ATNF), or if only available in other bands, extrapolate to 600\,MHz using an assumed spectral index value of $-$1.8. Using the minimum detectable flux density and the previously recorded flux density, we can predict a signal-to-noise ratio:
\begin{equation}
    \left(\frac{S}{N}\right)_{predicted} = \frac{S_{600}}{S_{min}}
\end{equation}
We compare this predicted signal-to-noise ratio to the actual signal-to-noise ratio of the pulse profile from folded CHAMPSS observations, as seen in Figure \ref{fig:sensitivity}. The pulse profile is created by phase connecting multiple days (up to 30 days) of observations using the known pulsar ephemeris. In general, we find good agreement between the predicted and detected signal-to-noise ratio.

\vspace{0.5cm}
\subsection{Cygnus Region / Implications for Galactic electron models}
\label{sec:cygnus}

\begin{figure}
    \centering
    \includegraphics[width=1.0\columnwidth]{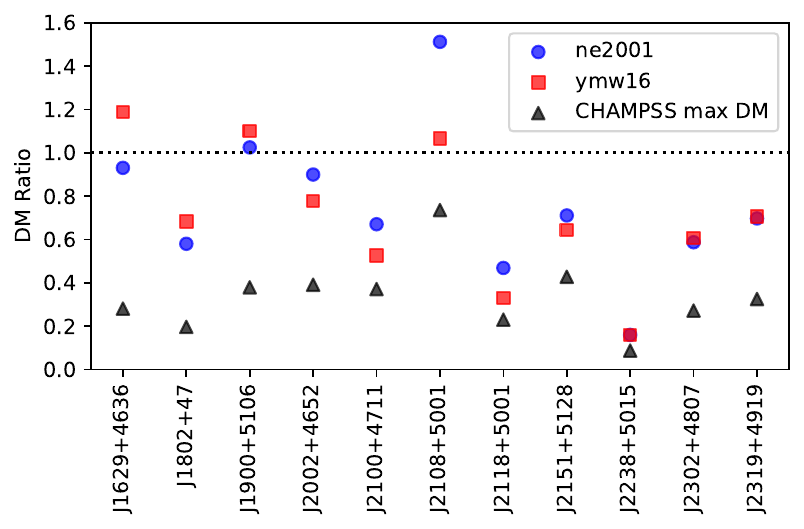} \\
    \caption{Ratio of the new discovered pulsar's DM to the maximum from NE2001, YMW16, and the max search DM from our pointing map. Although Galactic, three pulsars are in excess of NE2001 and/or YMW16, indicating a higher DM than predicted for the entire Milky Way.  Despite the excess, all pulsars have DM well below the search limit of CHAMPSS. }
    \label{fig:DMprofs}
\end{figure}

Three of the newly discovered pulsars show DM in excess of the maximum predicted line-of-sight DM 
from one, or both of the NE2001, YMW16 galactic electron density models (see Figure \ref{fig:DMprofs}).  Two pulsars, PSR~J1629+4636 and PSR~J1900+5106, are well off the Galactic Plane (with $g_{b} = 43.1^{\circ}, 19.4^{\circ}$, respectively), in regions devoid of many pulsars.  Such sightlines are poorly sampled for Galactic electron density models, and thus more uncertain.  

\begin{figure}
    \centering
    \includegraphics[width=1.0\columnwidth, trim={0 0 0 0.6cm}, clip]{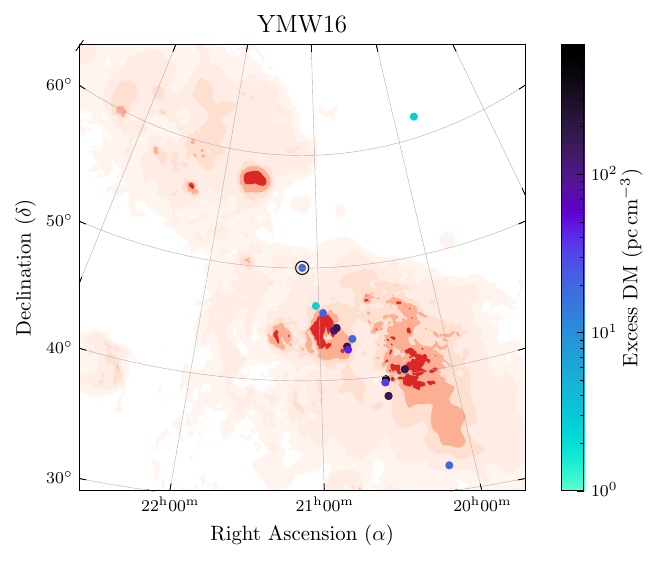}
    \caption{Excess DM beyond the Galactic maximum predicted by the YMW16 model in the vicinity of PSR~J2108+5001. Pulsars with DMs within model predictions are not shown and PSR~J2108+5001 is highlighted with a black ring. Contours show the \citet{finkbeinerFullSkyHaTemplate2003} $\mathrm{H}\alpha$ map at values of 10, 20, 50, 100, 200, and 500 Rayleigh, and the pulsar colors show the DM in excess of YMW16.}
    \label{fig:excess_dm_j2108_halpha}
\end{figure}

The third pulsar is PSR~J2108+5001, which is in the Galactic Plane at $g_{l} = 91.2^{\circ}, g_{b} = 1.47^{\circ}$.  
This lies on the outskirts of the Cygnus star-forming region, seen from recent pulsar discoveries by FAST to have DM far in excess of existing Galactic DM models \citep{han+21}.  \citet{ocker+24} show how many pulsars with excess DM and scattering intersect HII regions, including the aforementioned sources. We overlay all pulsars in excess of the predicted YMW16 dispersion measure onto H-$\alpha$ contours of the \citet{finkbeinerFullSkyHaTemplate2003} full-sky map, shown in Figure \ref{fig:excess_dm_j2108_halpha}. J2108+5001 is near the edge of a large region of higher H$\alpha$ emission which spans $\sim 20^{\circ}$ and includes the FAST-GPPS pulsars. However, this larger structure could just be due to the Orion-Cygnus arm of the Galaxy. The line of sight to J2108+5001 also intersects a known HII region, G090.856+01.691~\citep{andersonWISECatalogGalactic2014}, which could explain its excess DM. In either case, it is clear that the Cygnus region is not well modelled by either NE2001 or YMW16.

\begin{figure}
    \centering
    \includegraphics[width=1.0\columnwidth]{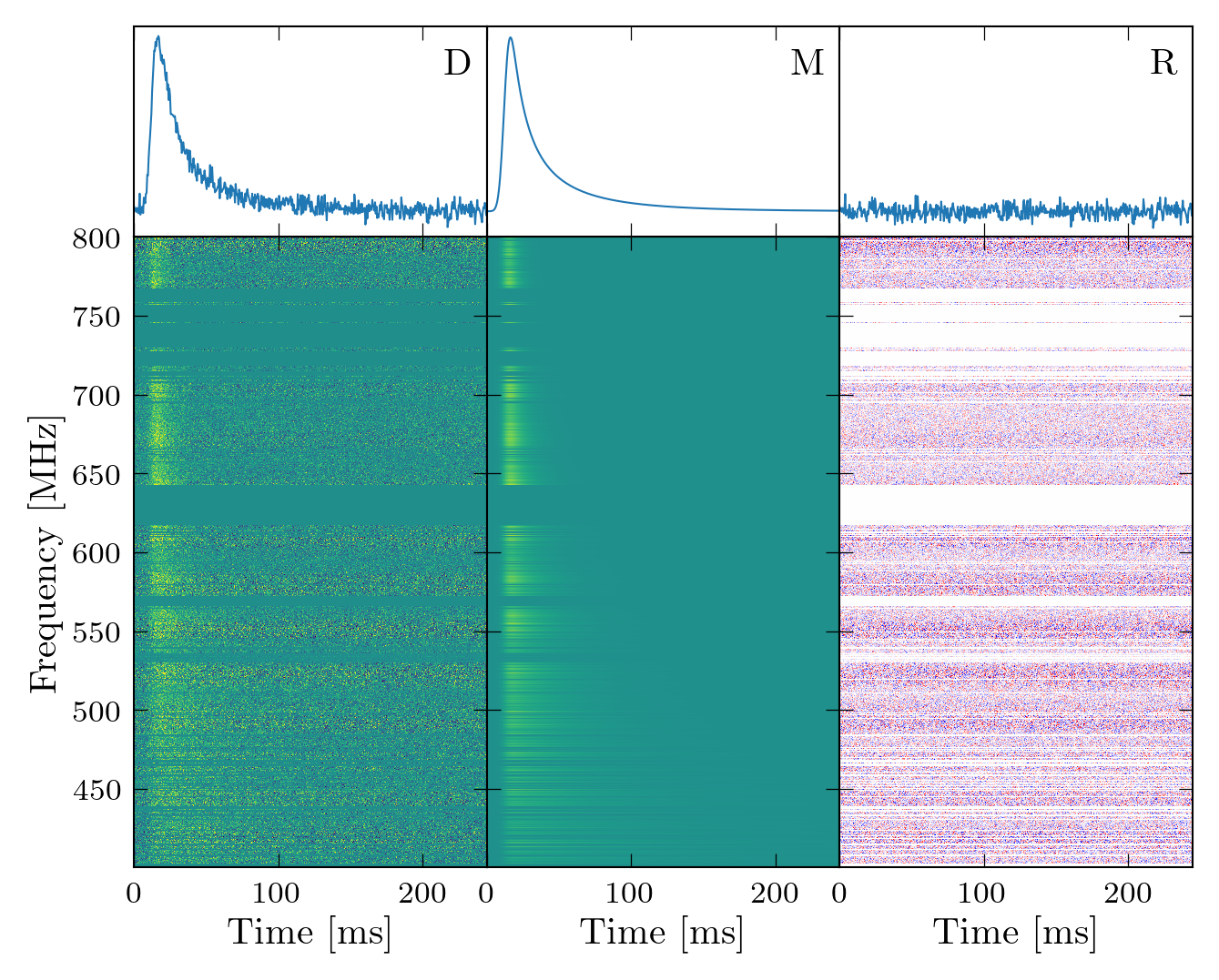}
    \caption{Fitburst~\citep{fonsecaModelingMorphologyFast2024} modelling of the pulse profile for PSR~J2108+5001. The central plot shows the model, with the data to the left and residuals to the right.}
    \label{fig:j2108_fitburst}
\end{figure}

J2108+5001 is also noticeably scattered.  We fit a scattering time using \texttt{fitburst} \citep{fonsecaModelingMorphologyFast2024}, after stacking the profiles of 10 days together for increased S/N.  The resulting scattering time is $\tau = 14.7\pm0.3\,$ms referenced to $600\,$MHz, or $\tau = 1.79\pm0.5\,$ms at 1\,GHz.  The profile and fit results are shown in Figure \ref{fig:j2108_fitburst}.

While not extraordinary compared to the most scattered pulsars known, such scattering times have consequences for other transient searches. CHAMPSS will struggle to discover more rapidly rotating pulsars in this region, as the time delay will be comparable or greater to the rotational period.  CHIME/FRB has significantly reduced sensitivity to bursts with widths $\gtrsim 10\,$ms \citep{merryfield+23}, suggesting it may be difficult to detect an FRB through this line of sight, or any others intersecting prominent HII regions.

\section{Conclusions and Future Work}

We have presented an overview of CHAMPSS, a periodicity search using the CHIME/FRB data stream which will search for pulsars daily, and in long-term power spectra stacks.  The pipeline runs in real time, and in a commissioning survey covering $\approx 6\%$ of the sky over $2\,$months, we discovered 11 new pulsars in the period range $0.2 - 1.5$\,seconds.  

CHAMPSS will scale up in 2025, following significant hardware upgrades, and software optimizations.  When operating at scale, we plan to process $>1/4$ of the sky in real time, with the aim of surveying the full northern sky.  This survey will be complementary to sensitive targeted surveys; repeated pointing will allow us the ability to discover many intermittent sources which could otherwise be missed in a single pointing (whether intrinsically intermittent, or through propagation effects such as scintillation, lensing, eclipsing).  Discoveries of more pulsars on under-searched lines of sight will help better constrain Galactic electron models, additionally useful in determining the boundary between high-DM Galactic pulsars, and low-DM FRBs (e.g. \citealt{cook+23}).  Even with only a few discoveries, it is clear that CHAMPSS will be fruitful in this regard, with three sources having DM in excess of the existing models.

As the survey progresses, we will include extensions to the standard FFT-based periodicity search.  One such search is a Fast Folding Algorithm (FFA, \citealt{staelin69}), which coherently sums all harmonics and is particularly well suited to long-pulsars which tend to have short duty cycles (e.g. \citealt{parent+18}).  While FFAs are computationally expensive over equivalent search space of a power spectrum search, the trial space is small if the search is restricted to long periods.  Additionally, we will investigate the efficacy of searching power spectra stacks for the change in observed frequency caused by the orbital response.  These extensions will be expanded on in future work.

One major hurdle of CHAMPSS is dealing in real time with such a large volume of candidates - for the scale of this survey this is far in excess of a few humans verifying candidates.  We are developing a machine learning algorithm to filter candidates, trained on real pulsars and realistic injected pulsar signals.  These additional features will be presented in future work.

CHAMPSS represents a new era of pulsar search - as modern radio observatories move to large-N, small-D designs, and have PB/day data volumes.  Similar survey designs can be used for e.g. CHORD, and the upcoming SKA.

\section*{Acknowledgements}

We acknowledge that CHIME is located on the traditional, ancestral, and unceded territory of the Syilx/Okanagan people. We are grateful to the staff of the Dominion Radio Astrophysical Observatory, which is operated by the National Research Council of Canada.  

CHIME is funded by a grant from the Canada Foundation for Innovation (CFI) 2012 Leading Edge Fund (Project 31170) and by contributions from the provinces of British Columbia, Qu\'{e}bec and Ontario. The CHIME/FRB Project, which enabled development in common with the CHIME/Pulsar instrument, is funded by a grant from the CFI 2015 Innovation Fund (Project 33213) and by contributions from the provinces of British Columbia and Qu\'{e}bec, and by the Dunlap Institute for Astronomy and Astrophysics at the University of Toronto. Additional support was provided by the Canadian Institute for Advanced Research (CIFAR), McGill University and the McGill Space Institute thanks to the Trottier Family Foundation, and the University of British Columbia. The CHIME/Pulsar instrument hardware was funded by NSERC RTI-1 grant EQPEQ 458893-2014.

CHAMPSS acknowledges funding from the Canadian Initiative for Radio Astronomy Data Analysis (CIRADA).
CIRADA is funded by a grant from the Canada Foundation for Innovation 2017 Innovation Fund (Project 35999) and by the Provinces of Ontario, British Columbia, Alberta, Manitoba and Quebec, in collaboration with the National Research Council of Canada, the US National Radio Astronomy Observatory and Australia’s Commonwealth Scientific and Industrial Research Organisation
This research was enabled in part by support provided by Calcul Québec, the BC Digital Research Infrastructure Group, and the Digital Research Alliance of Canada (alliancecan.ca).
This work was supported in part by a Canada Excellence Research Chair in Transient Astrophysics (CERC-2022-00009).

E.F. is supported by the National Science Foundation under grant AST-2407399.
K.W.M. holds the Adam J. Burgasser Chair in Astrophysics.
J.M.P. acknowledges the support of an NSERC Discovery Grant (RGPIN-2023-05373).
A.B.P. is a Banting Fellow, a McGill Space Institute~(MSI) Fellow, and a Fonds de Recherche du Quebec -- Nature et Technologies~(FRQNT) postdoctoral fellow.
U.P. is supported by the Natural Sciences and Engineering Research Council of Canada (NSERC) [funding reference number RGPIN-2019-06770, ALLRP 586559-23, RGPIN-2025-06396] , Ontario Research Fund – Research Excellence (ORF-RE Fund), Canadian Institute for Advanced Research (CIFAR), AMD AI Quantum Astro.
The National Radio Astronomy Observatory is a facility of the National Science Foundation operated under cooperative agreement by Associated Universities, Inc. SMR is a CIFAR Fellow and is supported by the NSF Physics Frontiers Center award 2020265.
Pulsar and FRB research at UBC is funded by an NSERC Discovery Grant and by the Canadian Institute for Advanced Research.

\bibliography{biblio.bib}{}
\bibliographystyle{aasjournal}



\section{Appendix}

\subsection{Description of Candidate Plots}
\label{sec:appendix_cand_plot}

\begin{figure*}
    \centering
    \includegraphics[width=0.95\textwidth]{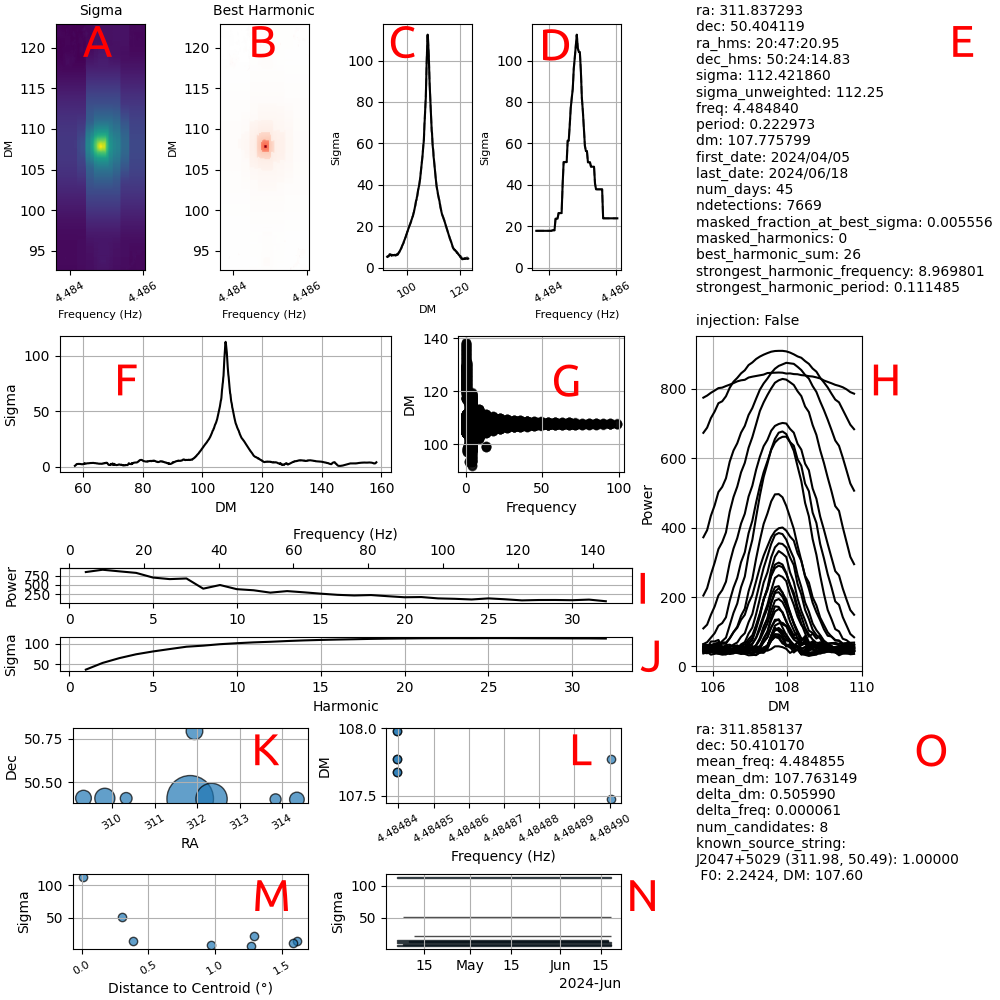}
    \caption{Example multi-pointing candidate plot of PSR~J2047+5029. Red alphabetic labels have been added to the individual segments of the plot.}
    \label{fig:example_mp_appendix}
\end{figure*}

Figure \ref{fig:example_mp_appendix} shows an exemplary multi-pointing candidate plot. This section describes the various segments of the candidate plot based on the read labels that were added to the plot.
Fields A-J show information about the strongest single-pointing candidate, while fields K-O show information about all single-pointing candidates included in the multi-pointing candidate.

\begin{enumerate}[label=\Alph*]
\item Sigma as a function of DM and frequency. The plotted value is the maximum of the different harmonic sums performed during the search.
\item The color shows the harmonic sum for each pixel in plot A with maximum sigma. A stronger red implies more summed harmonics. The alpha value of each pixel is controlled by the corresponding sigma value.
\item Slice through plot A along the DM axis.
\item Slice through plot A along the frequency axis.
\item Text field describing the parameters of the single-pointing candidate.
\begin{description}
    \item[ra] Right ascension in degrees.
    \item[dec] Declination in degrees.
    \item[ra\_hms] Right ascension in hh:mm:ss.s format.
    \item[dec\_hms] Right ascension in +dd:mm:ss format.
    \item[sigma] Maximum sigma of the candidate. Masked frequency bins are not included in the number of summands in the expected distribution during sigma calculation.
    \item[sigma\_unweighted] Maximum sigma of the candidate. Masked frequency bins are included in the number of summands in the expected distribution during sigma calculation.
    \item[freq] Best frequency of the candidate.
    \item[period] Best period of the candidate.
    \item[dm] Best DM of the candidate.
    \item[first\_date] First date included in the power spectrum stack.
    \item[last\_date] Last date included in the power spectrum stack.
    \item[ndays] Number of days included in the power spectrum stack.
    \item[detections] Number of detections that were clustered to form this candidate.
    \item[masked\_fraction\_at\_best\_sigma] Fraction of masked frequency bins at maximum sigma. This will also show if some frequency bins have been masked in individual days.
    \item[masked\_harmonics] Number of harmonics of best candidate which have been masked completely.
    \item[best\_harmonic\_sum] Number of summed harmonics which results in the highest sigma. Maximum position of plot J.
    \item[strongest\_harmonic\_frequency] Frequency of the harmonic with the most power.
    \item[strongest\_harmonic\_period] Period of the harmonic with the most power.
    \item[injection] Whether the candidate is the result of an injection or not.

\end{description}
\item Sigma as a function of DM along the best frequency. Same plot as C but with a bigger span.
\item Clustered detections included in this candidate.
\item Raw power at best frequency across the first 32 harmonics as a function of DM. When a harmonic is completely masked (as in the left plot of Figure \ref{fig:candidate-examples}) a red circle is drawn at that harmonic. This id done to more easily discern masked harmonics from weak harmonics.
\item Power of each harmonic at the best DM.
\item Sigma as a function of summed harmonics.
\item Sky positions of all single-pointing candidates. Size represents the sigma of the candidates.
\item Frequency and DM of all candidates.
\item Sigma as a function of the distance to centroid for all candidates.
\item Date span  and sigma of each candidate.
\item Text Text field describing the parameters of the multi-pointing candidate.
\begin{description}
    \item[ra] Right ascension in degrees of the centroid of the multi-pointing candidate. When calculating this centroid, sigma is used as the weight of each candidate. 
    \item[dec] Declination in degrees of the centroid of the multi-pointing candidate.
    \item[mean\_frequency] Mean frequency.
    \item[mean\_dm] Mean DM.
    \item[delta\_frequency] Frequency span.
    \item[delta\_DM] DM span.
    \item[num\_candidates] Number of single-pointing candidates that have been grouped to form this multi-pointing candidate.
    \item[known\_source\_string] Text field describing the output of the known source sifter. For each known source with a positive classification the name, right ascension, declination, likelihood that the candidate is the known source, frequency and DM are shown.
\end{description}

\subsection{Timing solution residuals}

\begin{figure}
    \centering
    \includegraphics[width=\linewidth]{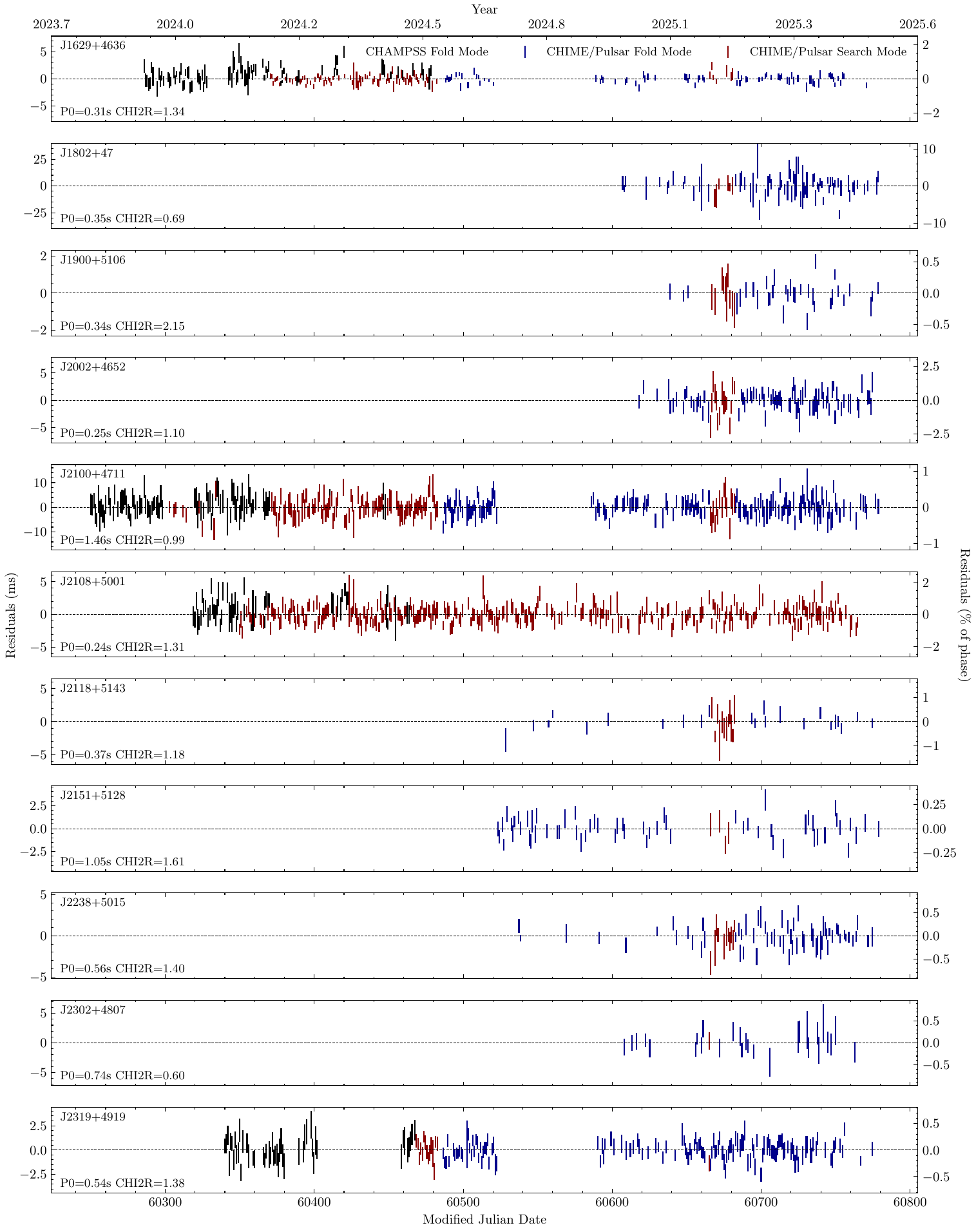}
    \caption{Timing residuals for 11 newly discovered pulsars. Pulsar names (top), periods, and reduced $\chi^2$ (bottom) of the fit are given to the left of each plot. Black, blue, and red data points are residuals (with their errorbars) of TOAs from CHAMPSS, CHIME/Pulsar fold-mode, and CHIME/Pulsar filterbank (search-mode) observations, respectively. 
    }
    \label{fig:timing_residuals}
\end{figure}

\end{enumerate}

\end{document}

%% file: authors.tex
\author[0009-0002-2429-3947]{Christopher Andrade}
  \affiliation{Department of Physics and Astronomy, University of British Columbia, 6224 Agricultural Road, Vancouver, BC V6T 1Z1 Canada}
\author[0000-0001-8537-9299]{P.~J.~Boyle}
  \affiliation{Department of Physics, McGill University, 3600 rue University, Montr\'eal, QC H3A 2T8, Canada}
\author[0000-0002-1800-8233]{Charanjot Brar}
  \affiliation{NRC Herzberg Astronomy and Astrophysics, 5071 West Saanich Road, Victoria, BC V9E2E7, Canada}
\author[0009-0007-0757-9800]{Alyssa Cassity}
  \affiliation{Department of Physics and Astronomy, University of British Columbia, 6224 Agricultural Road, Vancouver, BC V6T 1Z1 Canada}
\author[0000-0002-1529-5169]{Kathryn Crowter}
  \affiliation{Department of Physics and Astronomy, University of British Columbia, 6224 Agricultural Road, Vancouver, BC V6T 1Z1 Canada}
\author[0000-0003-2319-9676]{Davor Cubranic}
  \affiliation{AbCellera Biologics, 150 W 4th Ave, Vancouver, BC V5Y 1G6, Canada}
\author[0009-0004-5775-8821]{Abigail K.~Denney}
  \affiliation{David A. Dunlap Department of Astronomy and Astrophysics, 50 St. George Street, University of Toronto, ON M5S 3H4, Canada}
\author[0000-0003-4098-5222]{Fengqiu Adam Dong}
  \affiliation{National Radio Astronomy Observatory, 520 Edgemont Rd, Charlottesville, VA 22903, USA}
\author[0000-0001-8384-5049]{Emmanuel Fonseca}
  \affiliation{Department of Physics and Astronomy, West Virginia University, PO Box 6315, Morgantown, WV 26506, USA }
  \affiliation{Center for Gravitational Waves and Cosmology, West Virginia University, Chestnut Ridge Research Building, Morgantown, WV 26505, USA}
\author[0009-0002-0330-9188]{Ajay Kumar}
  \affiliation{National Centre for Radio Astrophysics, Post Bag 3, Ganeshkhind, Pune, 411007, India}
\author[0000-0003-4634-5453]{Lars K\"{u}nkel}
  \affiliation{Department of Physics and Astronomy, University of British Columbia, 6224 Agricultural Road, Vancouver, BC V6T 1Z1 Canada}
\author[0000-0001-5523-6051]{Magnus L'Argent}
  \affiliation{Department of Physics, McGill University, 3600 rue University, Montr\'eal, QC H3A 2T8, Canada}
\author[0000-0002-1172-0754]{Dustin Lang}
  \affiliation{Perimeter Institute of Theoretical Physics, 31 Caroline Street North, Waterloo, ON N2L 2Y5, Canada}
\author[0000-0002-7164-9507 ]{Robert A.~Main}
  \affiliation{Department of Physics, McGill University, 3600 rue University, Montr\'eal, QC H3A 2T8, Canada}
  \affiliation{Trottier Space Institute, McGill University, 3550 rue University, Montr\'eal, QC H3A 2A7, Canada}
\author[0000-0002-4279-6946]{Kiyoshi W.~Masui}
  \affiliation{MIT Kavli Institute for Astrophysics and Space Research, Massachusetts Institute of Technology, 77 Massachusetts Ave, Cambridge, MA 02139, USA}
  \affiliation{Department of Physics, Massachusetts Institute of Technology, 77 Massachusetts Ave, Cambridge, MA 02139, USA}
\author[0000-0001-5536-4635]{Sujay Mate}
  \affiliation{Raman Research Institute, C. V. Raman Avenue, Sadashivanagar, Bangalore, Karnataka - 560080, India}
  \affiliation{Department of Astronomy and Astrophysics, Tata Institute of Fundamental Research, Mumbai, 400005, India}
\author[0000-0002-0772-9326]{Juan Mena-Parra}
  \affiliation{Dunlap Institute for Astronomy and Astrophysics, 50 St. George Street, University of Toronto, ON M5S 3H4, Canada}
  \affiliation{David A. Dunlap Department of Astronomy and Astrophysics, 50 St. George Street, University of Toronto, ON M5S 3H4, Canada}
\author[0000-0001-8845-1225]{Bradley W.~Meyers}
  \affiliation{Australian SKA Regional Centre (AusSRC), Curtin University, Bentley WA 6102, Australia}
  \affiliation{International Centre for Radio Astronomy Research (ICRAR), Curtin University, Bentley WA 6102, Australia}
\author[0000-0002-3616-5160]{Cherry Ng}
  \affiliation{Laboratoire de Physique et Chimie de l'Environnement et de l'Espace - Universit\'e d'Orl\'eans/CNRS, 45071, Orl\'eans Cedex 02, France}
\author[0000-0002-8912-0732]{Aaron B.~Pearlman}
  \affiliation{Department of Physics, McGill University, 3600 rue University, Montr\'eal, QC H3A 2T8, Canada}
  \affiliation{Trottier Space Institute, McGill University, 3550 rue University, Montr\'eal, QC H3A 2A7, Canada}
  \affiliation{Banting Fellow}
  \affiliation{McGill Space Institute Fellow}
  \affiliation{FRQNT Postdoctoral Fellow}
\author[0000-0003-2155-9578]{Ue-Li Pen}
  \affiliation{Dunlap Institute for Astronomy and Astrophysics, 50 St. George Street, University of Toronto, ON M5S 3H4, Canada}
  \affiliation{Institute of Astronomy and Astrophysics, Academia Sinica, Astronomy-Mathematics Building, No. 1, Sec. 4, Roosevelt Road, Taipei 10617, Taiwan}
  \affiliation{Canadian Institute for Theoretical Astrophysics, 60 St.~George Street, Toronto, ON M5S 3H8, Canada}
  \affiliation{Canadian Institute for Advanced Research, 180 Dundas St West, Toronto, ON M5G 1Z8, Canada; }
  \affiliation{Perimeter Institute of Theoretical Physics, 31 Caroline Street North, Waterloo, ON N2L 2Y5, Canada}
\author[0000-0001-5799-9714]{Scott M.~Ransom}
  \affiliation{National Radio Astronomy Observatory, 520 Edgemont Rd, Charlottesville, VA 22903, USA}
\author{Alexander P.~Roman}
  \affiliation{Perimeter Institute of Theoretical Physics, 31 Caroline Street North, Waterloo, ON N2L 2Y5, Canada}
  \affiliation{Department of Physics and Astronomy, University of Waterloo, Waterloo, ON N2L 3G1, Canada}
\author[0000-0002-2088-3125]{Kendrick Smith}
  \affiliation{Perimeter Institute of Theoretical Physics, 31 Caroline Street North, Waterloo, ON N2L 2Y5, Canada}
\author[0000-0001-6748-5290.]{Reynier Squillace}
  \affiliation{Anton Pannekoek Institute for Astronomy, University of Amsterdam, Science Park 904, 1098 XH Amsterdam, The Netherlands}
  \affiliation{National Radio Astronomy Observatory, 520 Edgemont Rd, Charlottesville, VA 22903, USA}
\author[0000-0001-9784-8670]{Ingrid Stairs}
  \affiliation{Department of Physics and Astronomy, University of British Columbia, 6224 Agricultural Road, Vancouver, BC V6T 1Z1 Canada}
\author[0000-0001-7509-0117]{Chia Min Tan}
  \affiliation{International Centre for Radio Astronomy Research (ICRAR), Curtin University, Bentley WA 6102, Australia}
\author[0009-0009-5322-0932]{Laurent Tarabout}
  \affiliation{Department of Physics, McGill University, 3600 rue University, Montr\'eal, QC H3A 2T8, Canada}
\author[0009-0009-9343-4193]{Xia Wenke}
  \affiliation{Department of Physics, McGill University, 3600 rue University, Montr\'eal, QC H3A 2T8, Canada}
\author[0000-0002-7076-8643]{Tarik J.~Zegmott}
  \affiliation{Department of Physics, McGill University, 3600 rue University, Montr\'eal, QC H3A 2T8, Canada}
  \affiliation{Trottier Space Institute, McGill University, 3550 rue University, Montr\'eal, QC H3A 2A7, Canada}